\documentclass[aps,prl,floatfix, amsmath,twocolumn, showpacs,superscriptaddress]{revtex4-1}
\usepackage[pdftex]{graphicx}            
\usepackage{amsfonts}
\usepackage{amsmath}
\usepackage{enumitem}
\newcommand{\etal}{et al.~}

\newcommand{\cref}[1]{Chap.~\ref{#1}}  
\begin{document}

\title{\Large Chimera states in networks of phase oscillators:\\the case of two small populations}

\author{Mark J. Panaggio}
\email[email: ]{panaggio@rose-hulman.edu}
\affiliation{Mathematics Department, Rose-Hulman Institute of Technology, Terre Haute, Indiana 47803, USA}
\author{Daniel M. Abrams}
\affiliation{Department of Engineering Sciences and Applied Mathematics, Northwestern University, Evanston, Illinois 60208, USA}
\affiliation{Department of Physics and Astronomy, Northwestern University, Evanston, Illinois 60208, USA}
\affiliation{Northwestern Institute on Complex Systems, Northwestern University, Evanston, Illinois 60208, USA}
\author{Peter Ashwin}
\affiliation{Centre for Systems, Dynamics and Control, Harrison Building, Exeter EX4 4QF, UK}
\author{Carlo R. Laing}
\affiliation{Institute of Natural and Mathematical Sciences, Massey University, Private Bag 102-904 NSMC, Auckland, New Zealand}


\begin{abstract}
Chimera states are dynamical patterns in networks of coupled oscillators in which regions of synchronous and asynchronous oscillation coexist.  Although these states are typically observed in large ensembles of oscillators and analyzed in the continuum limit, chimeras may also occur in systems with finite (and small) numbers of oscillators. Focusing on networks of $2N$ phase oscillators that are organized in two groups, we find that chimera states, corresponding to attracting periodic orbits, appear with as few as two oscillators per group and demonstrate that for $N>2$ the bifurcations that create them are analogous to those observed in the continuum limit. These findings suggest that chimeras, which bear striking similarities to dynamical patterns in nature, are observable and robust in small networks that are relevant to a variety of real-world systems.
\end{abstract}
\pacs{05.45.Xt, 89.75.Kd} 
\maketitle





\section{Introduction}
Synchronization is an important feature of swarms of fireflies \cite{Buck1966}, pedestrians on a footbridge \cite{Strogatz2005}, systems of Josephson junctions \cite{Wiesenfeld1998}, the power grid \cite{Filatrella2008},  oscillatory chemical reactions \cite{Kuramoto2003_1} and cells in the heart and brain \cite{Peskin1975,Michaels1987,Crook1997}. Since the pioneering work of Winfree \cite{Winfree1967} and Kuramoto \cite{Kuramoto1975}, mathematical models of arrays of coupled oscillators have been used to gain insight into the origin of spontaneous synchronization in a variety of different settings \cite{Strogatz2000}. In addition to uniform synchronous and asynchronous oscillation, many networks of oscillators are known to exhibit a type of partial synchronization known as a ``chimera'' state \cite{Kuramoto2002,Panaggio2014_2}. 

Chimera states are spatio-temporal patterns in which regions of coherence and incoherence coexist \cite{Abrams2004}. These patterns have been reported in analysis and simulation of coupled oscillators with a variety of network topologies  \cite{Kuramoto2002, Abrams2004, Shima2004, Abrams2006,Abrams2008, Martens2010_1, Martens2010_2, Omelchenko2012, Zhu2012, Panaggio2013,Panaggio2014} and appear to be robust to a variety of perturbations \cite{Laing2009_1,Laing2009_2, Laing2012,Laing2012_2, Yao2013}. Recently, they were also observed in experiments with optical \cite{Hagerstrom2012}, chemical \cite{Tinsley2012, Nkomo2013}, mechanical \cite{Martens2013} and electrochemical oscillators \cite{Schmidt2014}. They bear a strong resemblence to patterns of electrical activity in the human brain \cite{Laing2011, Wimmer2014,Tognoli2014,Laing2001}. 

When they coined the term ``chimera'' state, Abrams and Strogatz defined it as a state in which an array of identical oscillators splits into domains of synchonized and desynchronized oscillation.  Recently Ashwin \etal formalised this idea into a definition of a ``weak chimera'' \cite{Ashwin2015} such that one can prove the existence of, and investigate the stability and bifurcation of chimera-like solutions in small networks. For our purposes, we will use the term ``chimera state'' to refer to a trajectory in which two or more oscillators are synchronized (in phase and frequency) and one or more oscillators drift in phase and frequency with respect to the synchronized group; these are weak chimeras with the additional restriction of phase synchronization.


In nature oscillators often experience non-local interactions with other oscillators that promote synchronization. It is natural to model the dynamics on these networks using finite systems of differential equations. In models like the Kuramoto model, fully synchronized states can be understood as stationary solutions in a rotating frame of reference.  Unfortunately, chimera states and other partially synchronized solutions are more difficult to characterize due to fluctuations in the local degree of synchrony, which is measured by an order parameter. As a result, until recently \cite{Ashwin2015,wolfrum2015}, little progress has been made in analyzing chimera states for finite networks.

Instead, theoretical investigations of chimera states often replace a finite network of oscillators (discrete) with an infinite network (continuum)~\cite{Abrams2004,Abrams2008,Martens2010_1,Omelchenko2013}. In the continuum limit, the order parameter is stationary for a variety of synchronized and desynchronized solutions, including chimera states. This makes it possible to characterize chimera states by solving an eigenvalue problem.  Unfortunately, rigorous analysis of the detailed dynamics and stability of chimera states remains difficult, even in the continuum limit (it is possible in certain lower dimensional cases \cite{Omelchenko2013}).  So, researchers typically discretize the theoretical solutions from the continuum limit and assess the stability using numerical simulations with large ensembles of oscillators. The implicit assumption is that these two systems, one with an infinite number of oscillators and the other with a finite number, should behave in similar ways. However, at this time there is no rigorous justification for why this would need to be the case.  Although there is evidence that this holds for certain coupling schemes \cite{Ashwin2015}, there are cases where chimera states are stable in the continuum limit but not when the number of oscillators is finite.  For example, Wolfrum and Omel'chenko~\cite{Wolfrum2011_2} study a ring of oscillators with phases $\theta_k$ described by 
\begin{equation}
\frac{d\theta_k}{d t}=\omega- \frac{1}{2R} \sum_{j=k-R}^{k+R} \sin (\theta_k-\theta_j +\alpha) 
\label{eq:WOring}
\end{equation}
for $k=1,2,\dots N$ where $N$ is the number of oscillators. For finite $N$, and $R$ proportional to $N$ they show that chimera states are not attracting, but instead appear as chaotic transients.  This means one can observe a period of chimera-like behavior for certain initial conditions but the duration of this period is finite. In large systems ($N>50$) the observation of the collapse of a chimera in simulation is extremely rare, but in small networks ($N<25$) the lifetime can be quite short. It is unknown to what degree these observations apply to other topologies. 

Here we explore a simple network with two groups of oscillators. We show that in this network, the analogues of chimera states need not be transients --- they exist as stable periodic orbits with as few as two oscillators per group. 

\section{Two groups of $N$ oscillators for reducing $N$}

\begin{figure}[ht]
\center
	\includegraphics[width=0.8\columnwidth,trim=0cm 3cm 0cm 3cm, clip=true]{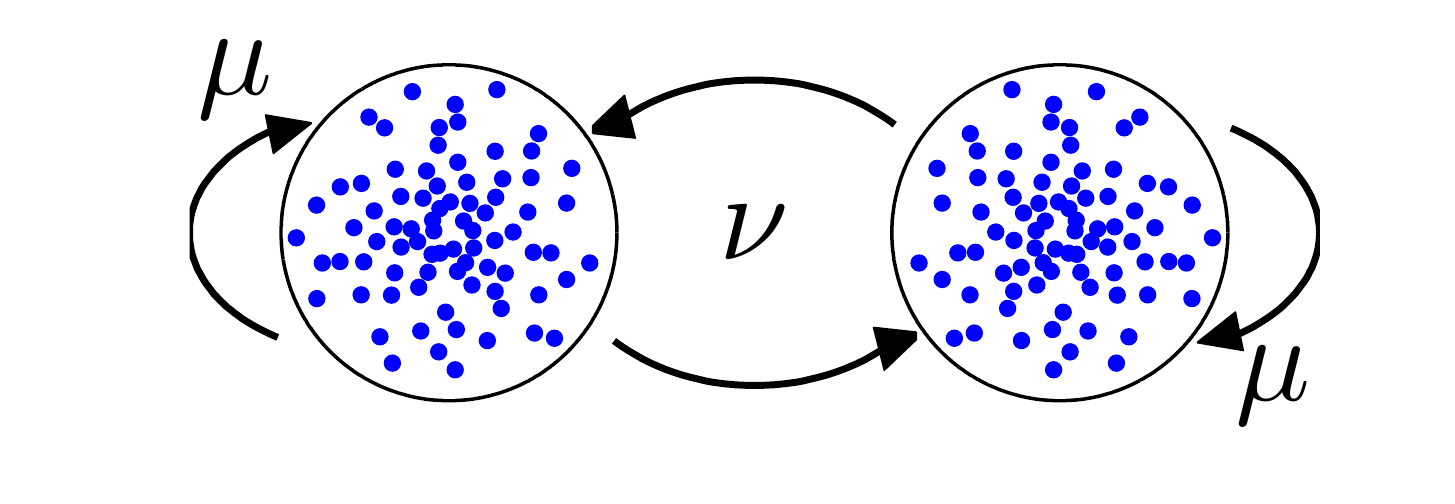} 
	\caption{Schematic diagram showing a network of $2N$ oscillators in two groups of $N$ coupled with strength $\mu$ within group and $\nu$ between groups.} 
	\label{fig:network2N}
\end{figure}

We start with the system studied in \cite{Abrams2008,Montbrio2004, Ott2008,Laing2012} consisting of two groups of $N$ phase oscillators with Kuramoto-Sakaguchi type coupling $\sin(\varphi+\alpha)=\cos(\varphi-\beta)$ for $\beta=\pi/2-\alpha$. We assume an inter-group coupling strength of $\nu=(1-A)/2$ and intra-group coupling $\mu=(1+A)/2$, where $0\leq A\leq 1$; figure~\ref{fig:network2N} illustrates such a network. Let $\{\theta_i\}_{i=1}^N$ and $\{\phi_i\}_{i=1}^N$ represent the phases of oscillators in groups 1 and 2 respectively.  If all oscillators have the same natural frequency $\omega$, then their phases are governed by 
\begin{align}
\frac{d\theta_i}{d t}&=\omega-\left(\frac{1+A}{2N}\right)\sum_{j=1}^N\cos(\theta_i-\theta_j-\beta)\nonumber\\
&-\left(\frac{1-A}{2N}\right)\sum_{j=1}^N\cos(\theta_i-\phi_j-\beta), \label{eq:dthdt} \\
\frac{d\phi_i}{d t}&=\omega -\left(\frac{1+A}{2N}\right)\sum_{j=1}^N\cos(\phi_i-\phi_j-\beta)\nonumber\\&-\left(\frac{1-A}{2N}\right)\sum_{j=1}^N\cos(\phi_i-\theta_j-\beta) \label{eq:dphidt}
\end{align}
Note that the $A$ and $\beta$ used here are the same as those in Abrams \etal \cite{Abrams2008}.

\subsection{Dynamics and bifurcations for $N=\infty$}

\begin{figure}[ht]
\center
	\includegraphics[width=0.8\columnwidth]{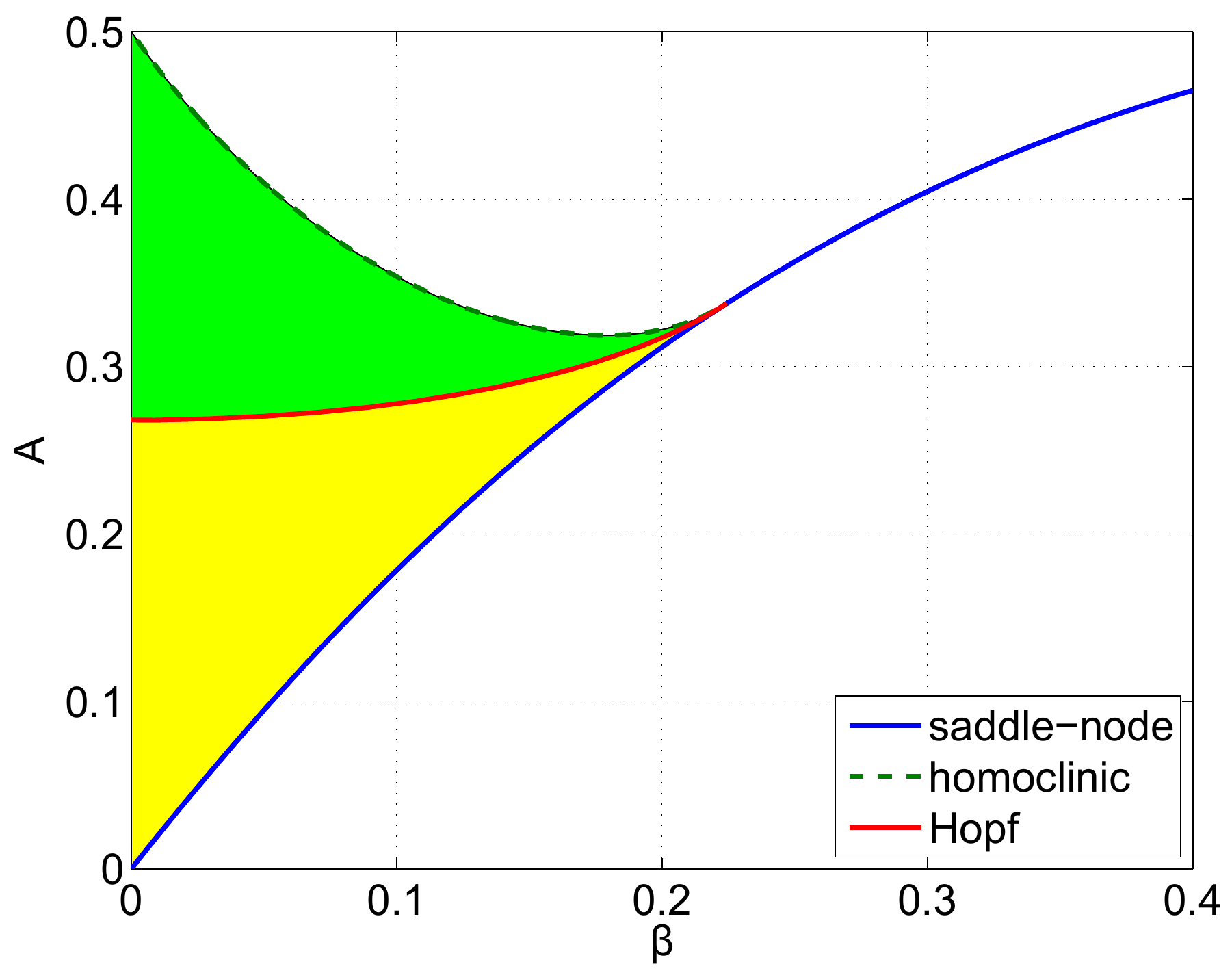} 
	\caption{Bifurcations of chimeras in the continuum limit $N=\infty$ of~\eqref{eq:dthdt}-\eqref{eq:dphidt}. (A similar figure appears in~\cite{Abrams2008}.) The yellow and green shaded regions correspond to the respective regions of existence for stationary and breathing chimeras. 
	} 
	\label{fig:danny08}
\end{figure}

The dynamics of~\eqref{eq:dthdt}-\eqref{eq:dphidt} were studied in the limit $N\rightarrow\infty$ in \cite{Abrams2008}. Summarizing, they use the Ott/Antonsen ansatz~\cite{Ott2008,Ott2009_1} to derive a set of ordinary differential equations (ODEs) satisfied by the complex order parameters for the two groups of oscillators (which measure the degree of synchrony within each group). A chimera state in our sense corresponds to one group being perfectly synchronized (so that the magnitude of its order parameter $\left|z\right| = \left| \frac{1}{N} \sum_{j=1}^N e^{i \theta_j}\right|$ is equal to $1$) while oscillators in the other group are asynchronous (so that magnitude of its order parameter satisfies $0\leq \left|z\right|<1$).
In this case, Abrams \etal derive a pair of real ODEs for the magnitude of the order parameter of
the asynchronous group and the difference in phases of the two order parameters. Analyzing these equations on varying $A$ and $\beta$, they obtain the behavior shown in Fig.~\ref{fig:danny08}.
For small $\beta$, as $A$ is increased from zero, two chimera states (one stable and one a saddle) are created in a saddle-node bifurcation. As $A$ increases the stable chimera undergoes a supercritical Hopf bifurcation, leading to a ``breathing'' chimera. Increasing $A$ further results in this solution colliding with the saddle chimera in a homoclinic bifurcation, and no stable chimeras remain.

 


\subsection{Dynamics for finite $N$ }

\begin{figure}[ht!]
\includegraphics[width=0.9\columnwidth]{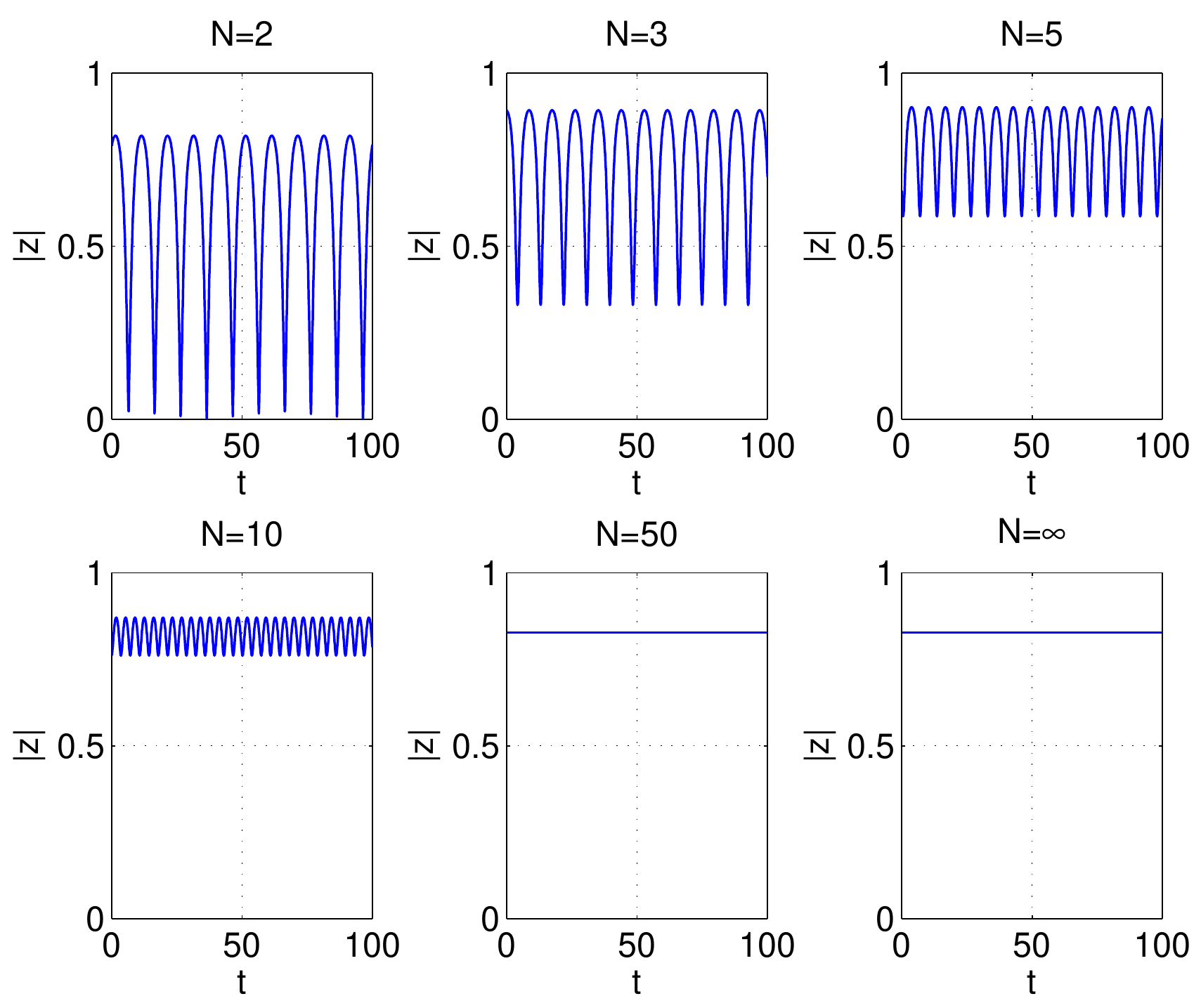}
\caption{The trajectories of the order parameter in the asynchronous group for various values of $N$. Parameters: $A=0.1,\beta=0.025$.}
\label{fig:trajNa}
\end{figure}

Generating initial conditions consistent with the stationary chimera in the continuum limit for varying $N$ we obtain the trajectories for order parameter $z=\frac{1}{N}\sum_{j=1}^N e^{i\theta_j}$  shown in Fig.~\ref{fig:trajNa}.
When $N$ is small, it appears that the dynamics of the order parameter in the desynchronized group are dominated by fluctuations, a finite size effect not observed in the infinite $N$ limit. As $N$ increases, the magnitude of these fluctuations decreases and the dynamics approach those of the continuum limit. For $N>50$ the dynamics of the order parameter are virtually indistinguishable from the results predicted by the continuum theory.

\begin{figure}[ht!]
\includegraphics[width=0.9\columnwidth]{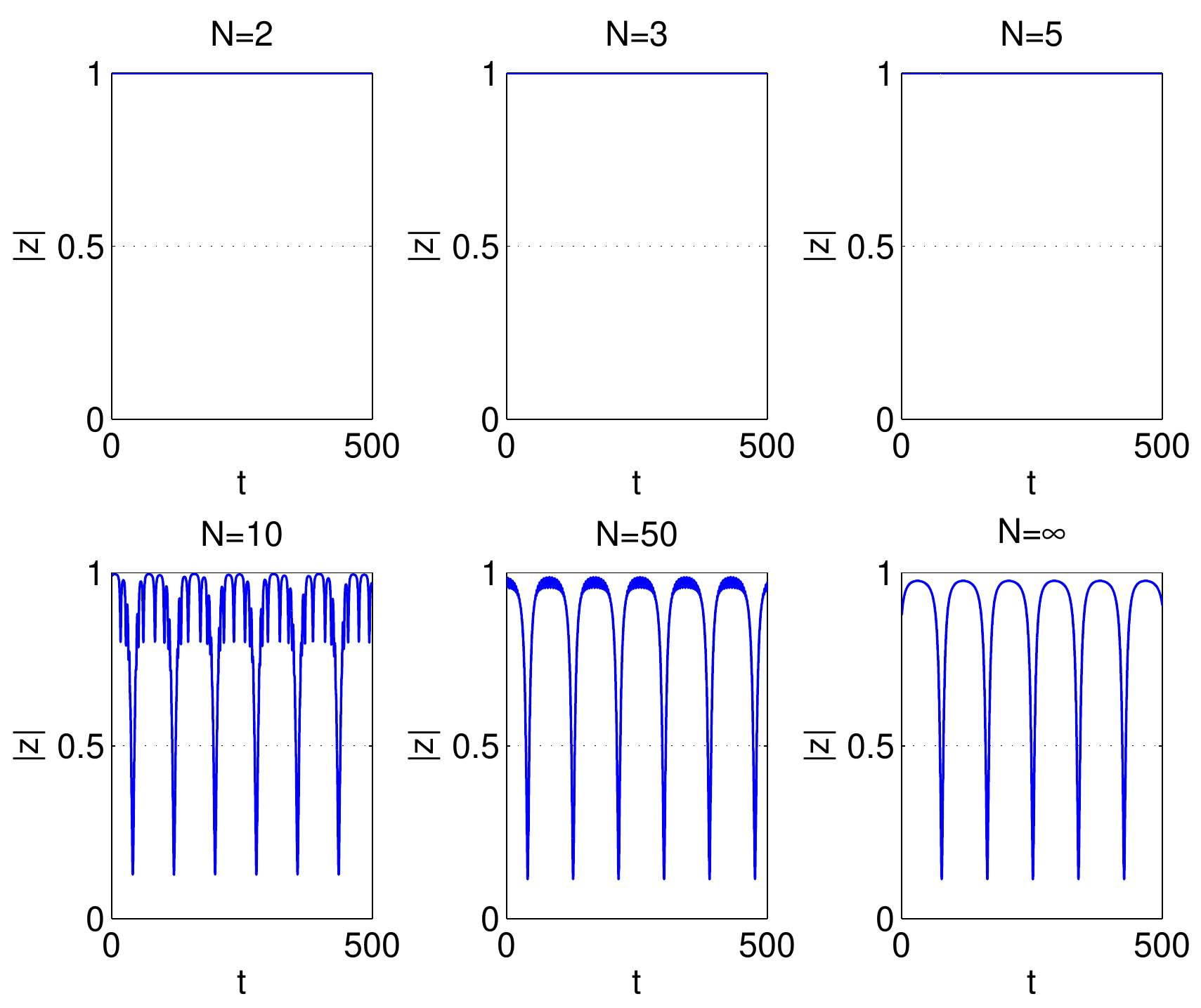}
\caption{The trajectories of the order parameter in the asynchronous group for various values of $N$. Parameters: $A=0.4,\beta=0.025$.}
\label{fig:trajNb}
\end{figure}

For parameter values above the Hopf bifurcation shown in Fig.~\ref{fig:danny08} we obtain the results shown in Fig.~\ref{fig:trajNb}. For large but finite $N$ we see similar fluctuations superimposed on the oscillations of the order parameter shown in the $N=\infty$ case. However, such an oscillatory state does not appear to be stable when $N$ is small, and it is not clear why this is the case. Therefore, at least for small $N$, the finite size effects dominate the dynamics and the intuition that one can gain from the $N=\infty$ case is limited.

In order to study these phenomena further it is helpful to reduce the dimensionality of the system. Watanabe and Strogatz showed that arrays with $N$ globally-coupled identical phase oscillators possess $N-3$ constants of motion, provided that $N>3$, for almost all initial conditions~\cite{Watanabe1994}. In other words, only three governing equations are needed to describe the dynamics.  Pikovsky and Rosenblum~\cite{Pikovsky2008} extend this analysis to networks with multiple groups ~\cite{Pikovsky2008}. We therefore consider their alternative parametrization of the system in which the dynamics of each group are governed by
\begin{align}
  \frac{d\rho_j}{dt} & = \frac{1-\rho_j^2}{2}\mathrm{Re}{(Z_je^{-i\Phi_j})} \\
  \frac{d\Psi_j}{dt} & = \frac{1-\rho_j^2}{2\rho_j}\mathrm{Im}{(Z_je^{-i\Phi_j})} \\
  \frac{d\Phi_j}{dt} & = \omega+\frac{1+\rho_j^2}{2\rho_j}\mathrm{Im}{(Z_je^{-i\Phi_j})}
\end{align}
for $j=1,2$, where
\begin{align}
   Z_1= & \frac{-i(1+A)e^{i\beta}\rho_1 e^{i\Phi_1}\gamma_1-i(1-A)e^{i\beta}\rho_2 e^{i\Phi_2}\gamma_2}{2} \\
   Z_2= & \frac{-i(1+A)e^{i\beta}\rho_2 e^{i\Phi_2}\gamma_2-i(1-A)e^{i\beta}\rho_1 e^{i\Phi_1}\gamma_1}{2}
\end{align}
where
\[
   \gamma_j=\frac{1}{N\rho_j}\sum_{k=1}^N\frac{\rho_j+e^{i(\psi_k^{(j)}-\Psi_j)}}{1+\rho_je^{i(\psi_k^{(j)}-\Psi_j)}}
\]
and the $\psi_k^{(j)}$, with $k=1,\dots N$, are the constants associated with the Watanabe and Strogatz transformation for population $j$.
In this system, $\rho_j$ measures the degree of synchrony in group $j$ (but it is not equivalent to the order parameter in Ref.~\citep{Abrams2008}). $\Phi_j$ and $\Psi_j$ are related to the mean phase and spread of the phases of oscillators, respectively, within group $j$. Assume we are in a chimera state where $\rho_1=1$. Then $\gamma_1=1$ and
\begin{align}
   Z_1= & \frac{-i(1+A)e^{i\beta}e^{i\Phi_1}-i(1-A)\Gamma e^{i\xi}e^{i\beta}e^{i\Phi_2}}{2} \\
   Z_2= & \frac{-i(1+A)\Gamma e^{i\xi}e^{i\beta}e^{i\Phi_2}-i(1-A)e^{i\beta}e^{i\Phi_1}}{2}
\end{align}
where
\[
   \Gamma e^{i\xi}:= \rho_2\gamma_2=\frac{1}{N}\sum_{k=1}^N\frac{\rho_2e^{i\Psi_2}+e^{i\psi_k^{(2)}}}{e^{i\Psi_2}+\rho_2e^{i\psi_k^{(2)}}}
\]
Defining $\Delta=\Phi_1-\Phi_2$ we have
\begin{widetext}
\begin{align}
   \frac{d\rho_2}{dt} & = \left(\frac{1-\rho_2^2}{4}\right)\left[(1+A)\Gamma\sin{(\xi+\beta)}
   +(1-A)\sin{(\Delta+\beta)}\right] \label{eq:drhodt} \\
   \frac{d\Delta}{dt} & = \frac{1+A}{2}\left[-\cos{\beta}+\Gamma\left(\frac{1+\rho_2^2}{2\rho_2}\right)\cos{(\xi+\beta)}\right]
   +\frac{1-A}{2}\left[-\Gamma\cos{(\xi-\Delta+\beta)}+\left(\frac{1+\rho_2^2}{2\rho_2}\right)\cos{(\Delta+\beta)}\right] \label{eq:ddeltadt} \\
   \frac{d\Psi_2}{dt} & = -\left(\frac{1-\rho_2^2}{4\rho_2}\right)[(1+A)\Gamma\cos{(\xi+\beta)}+(1-A)\cos{(\Delta+\beta)}] \label{eq:dpsidt}
\end{align}
\end{widetext}
where $\Gamma$ and $\xi$ are functions of $\Psi_2$ and $\rho_2$. Now for a uniform distribution of the 
$\psi_k^{(2)}$, i.e.~$\psi_k^{(2)}=2\pi k/N$ (which is the appropriate choice for comparison with dynamics on the Ott/Antonsen manifold in the $N\rightarrow\infty$ case)  Pikovsky and Rosenblum~\cite{Pikovsky2008} showed that
\begin{align*}
   \gamma_2&=
   1+\frac{(1-\rho_2^2)(-\rho_2e^{-i\Psi_2})^N}{1-(-\rho_2 e^{-i\Psi_2})^N}.
\end{align*}
As $N\rightarrow\infty$, we have $\gamma_2\rightarrow 1, \xi\rightarrow 0, \Gamma\rightarrow \rho_2$ 
and~\eqref{eq:drhodt}-\eqref{eq:ddeltadt} decouple from~\eqref{eq:dpsidt}, resulting in equations~(12) from~\cite{Abrams2008} (after a redefinition of parameters).
For finite $N$ the equations are coupled. Simulating~\eqref{eq:drhodt}-\eqref{eq:dpsidt} with appropriate initial conditions for varying $N$ we obtain similar results to those in Figs.~\ref{fig:trajNa} and~\ref{fig:trajNb} (results not shown) although as noted above, $\rho_2$ does not correspond exactly to the magnitude of the order parameter plotted there. In all of these chimera states observed, $\Psi_2$ decreases monotonically.

\subsection{Dynamics and bifurcations for two groups of $N=4$ phase oscillators}

\begin{figure}[hb]
\includegraphics[scale=0.4,clip=true,trim= 0cm 13.6cm 0cm 0cm]{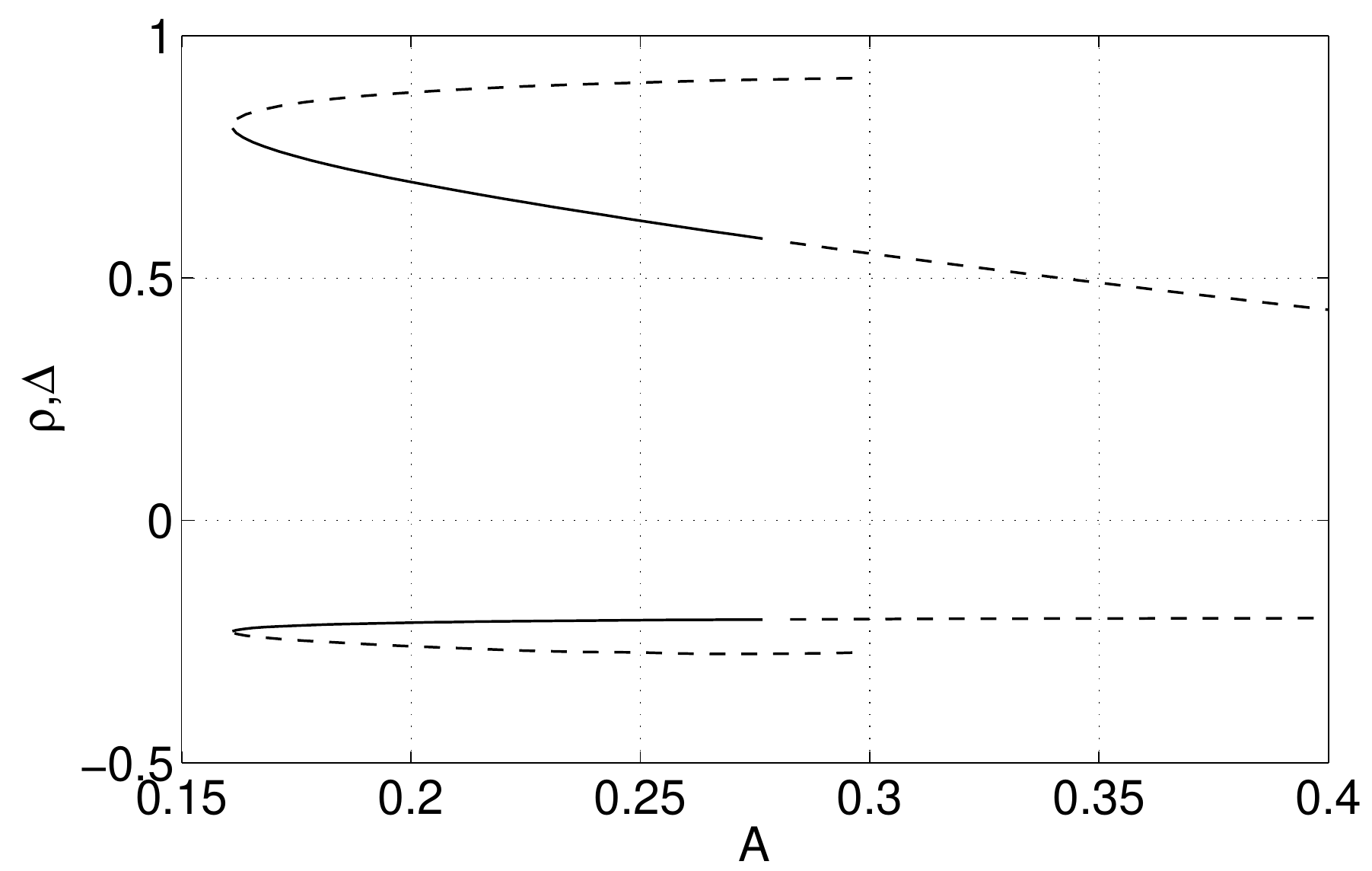}
\includegraphics[scale=0.4,clip=true,trim= 0cm 0cm 0cm 13.4cm]{fig5_ws4-eps-converted-to.pdf}
\caption{Top: the value of $\rho$ (positive) and $\Delta$ (negative) 
on the Poincar{\'e} section $\Psi=\pi$ for eqns.~\eqref{eq:drhodt}-\eqref{eq:dpsidt}. Solid: stable, dashed: unstable.
Bottom: period of periodic orbits. (Further continuation of the higher-period branch was not possible for numerical reasons.)
Parameters: $\beta=0.1,N=4$.}
\label{fig:ws4}
\end{figure}

We now consider the case $N=4$ in more detail.  To understand the dynamics we place a Poincar{\'e} section, $\Sigma$, in the flow at $\Psi\bmod 2\pi=\pi$ and record the values of $\rho$ and $\Delta$ as $\Psi$ decreases through $\Sigma$. We drop subscripts for notational convenience and seek chimera states corresponding to periodic solutions of~\eqref{eq:drhodt}-\eqref{eq:dpsidt}. These correspond to fixed points of the first return map for $\Sigma$. Following this fixed point as $A$ is varied and $\beta=0.1$ we use numerical continuation to obtain the results shown in Fig.~\ref{fig:ws4}. 

\begin{figure}[t]
\includegraphics[scale=0.4]{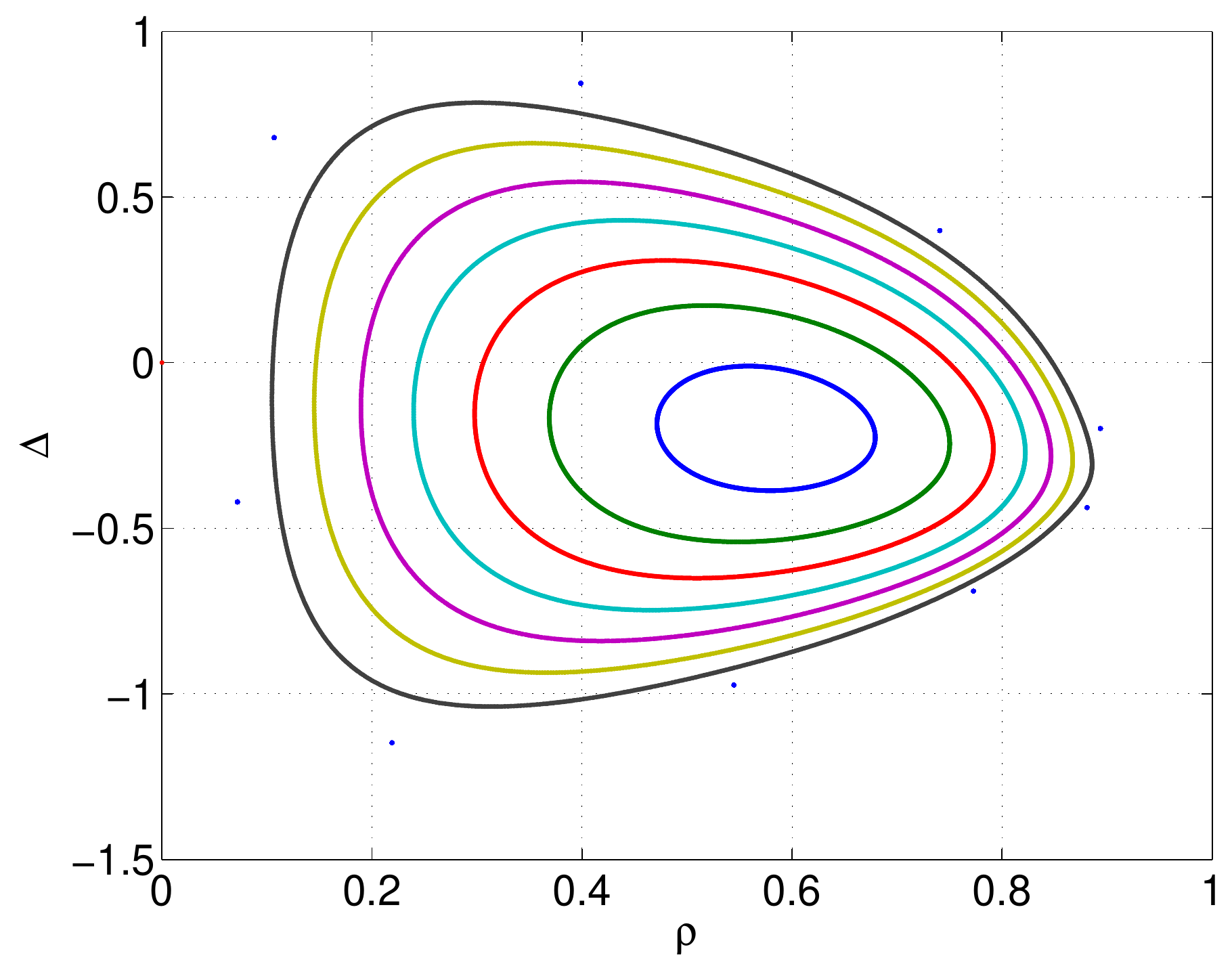}
\caption{The values of $\rho$ and $\Delta$ 
on the Poincar{\'e} section $\Psi=\pi$ for $A=0.28$ to $A=0.35$ (inner to outer) in steps of $0.01$.
Parameters: $\beta=0.1,N=4$.}
\label{fig:wsA}
\end{figure}

\begin{figure}[t]
\includegraphics[scale=0.4]{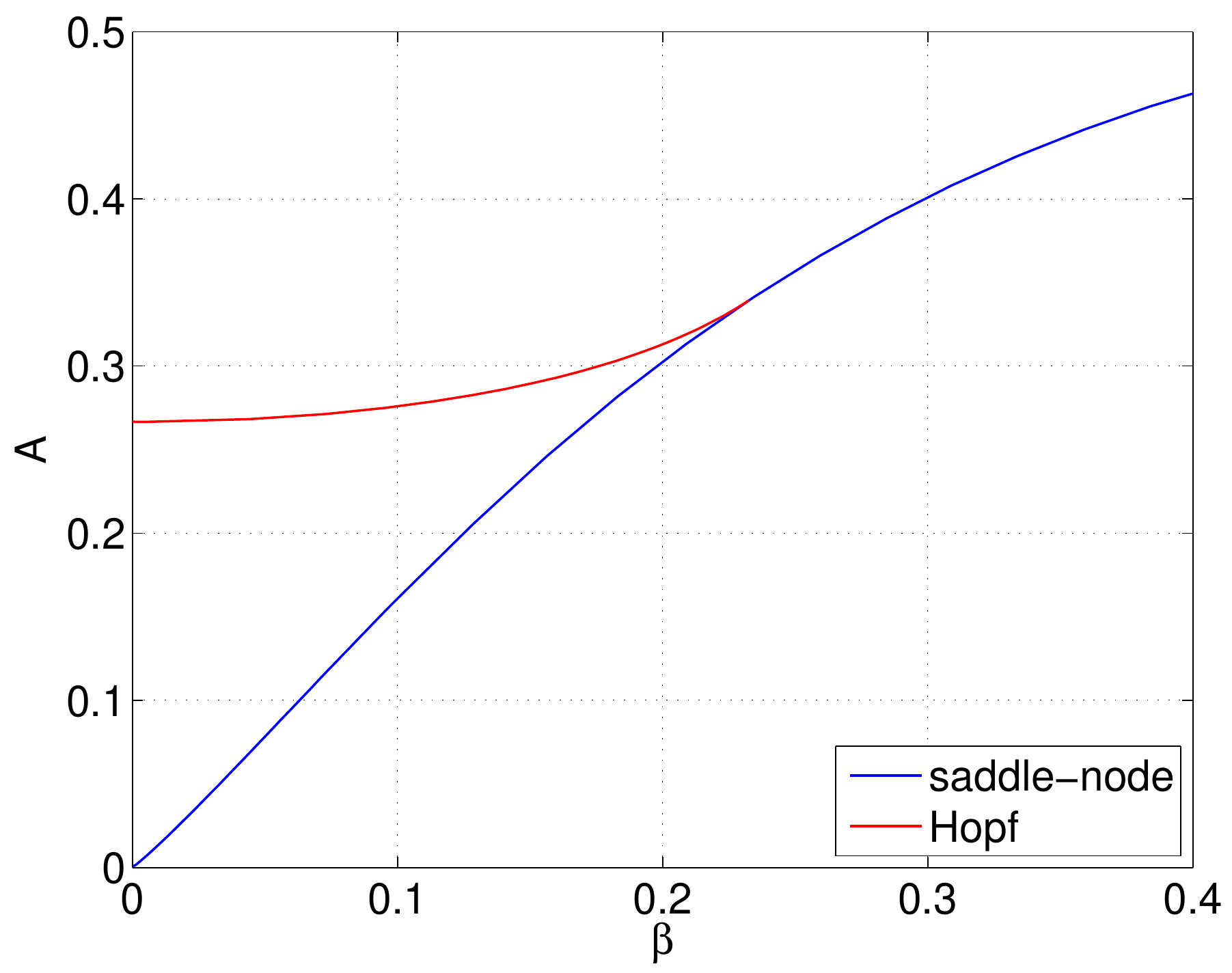}
\caption{Saddle-node and Hopf bifurcation curves for $N=4$. (See Fig.~\ref{fig:ws4}.)
}
\label{fig:wsA4}
\end{figure}

As $A$ is increased from zero a stable and unstable chimera are created in a saddle-node bifurcation. Increasing $A$ further results in the stable chimera becoming unstable via a Hopf bifurcation. Integrating~\eqref{eq:drhodt}-\eqref{eq:dpsidt} for values of $A$ above the Hopf bifurcation we obtain the results shown in Fig.~\ref{fig:wsA}. This strongly suggests that the Hopf bifurcation is supercritical, and that the stable solution created at this bifurcation is destroyed in a global bifurcation as $A$ is increased even further. This sequence of bifurcations is remarkably similar to the one observed in the infinite $N$ limit, and following the saddle-node and Hopf bifurcations shown in Fig.~\ref{fig:ws4} we obtain Fig.~\ref{fig:wsA4}, which should be compared with Fig.~\ref{fig:danny08}.  


Investigations for larger $N$ (results not shown) suggest that as $N\rightarrow\infty$ the curves of local bifurcations in Fig.~\ref{fig:wsA4} smoothly deform into the corresponding curves in Fig.~\ref{fig:danny08}. However, the stability of solutions indicated in Fig.~\ref{fig:ws4} is with respect to perturbations of just the variables in~\eqref{eq:drhodt}-\eqref{eq:dpsidt}, not to those in the original system~\eqref{eq:dthdt}-\eqref{eq:dphidt}. Simulations of~\eqref{eq:dthdt}-\eqref{eq:dphidt} for $N=4$ (again, not shown) suggest that chimera solutions shown as stable in  Fig.~\ref{fig:ws4} do correspond to stable chimera states in~\eqref{eq:dthdt}-\eqref{eq:dphidt}.

\subsection{Dynamics and bifurcations for two groups of $N=3$ phase oscillators}

When $N\leq 3$, the Watanabe-Strogatz ansatz used above no longer applies. Instead, we consider the original system~\eqref{eq:dthdt}-\eqref{eq:dphidt}. In a chimera state, one group is completely synchronized while the other is not, so, without loss of generality, we look for solutions in which the second group is synchronized, i.e.~$\phi_i=\phi$ for $i=1,\dots N$  and consider the phase differences $\psi_i=\theta_i-\phi$. In this rotating frame of reference, the dynamics of the desynchronized cluster decouple from the synchronized cluster and the phase differences $\psi_i$ satisfy
\begin{align}
\label{sys_rot}
\frac{d\psi_i}{dt}&= \frac{1+A}{2}\left[\cos{\beta}-\frac{1}{N}\sum_{j=1}^N\cos{(\psi_i-\psi_j-\beta)}\right] \nonumber \\
   &+\frac{1-A}{2}\left[\frac{1}{N}\sum_{j=1}^N\cos{(\psi_j+\beta)}-\cos{(\psi_i-\beta)}\right] 
\end{align}
for $i=1,2\dots N$. Note that, in general, this frame of reference does not have a constant frequency, because  $\phi$ satisfies
\begin{align}
\frac{d\phi}{dt}=\omega-\frac{1+A}{2}\cos{\beta}-\frac{1-A}{2N}\sum_{j=1}^N\cos(\psi_j+\beta).
\end{align}
Nonetheless, for the purpose of investigating the existence of chimera states, this equation can be ignored (in Appendix \ref{App:Stab} we derive conditions under which having a synchronized group is stable). 

\begin{figure}[ht]
\includegraphics[scale=0.4,clip=true,trim= 0cm 13.6cm 0cm 0cm]{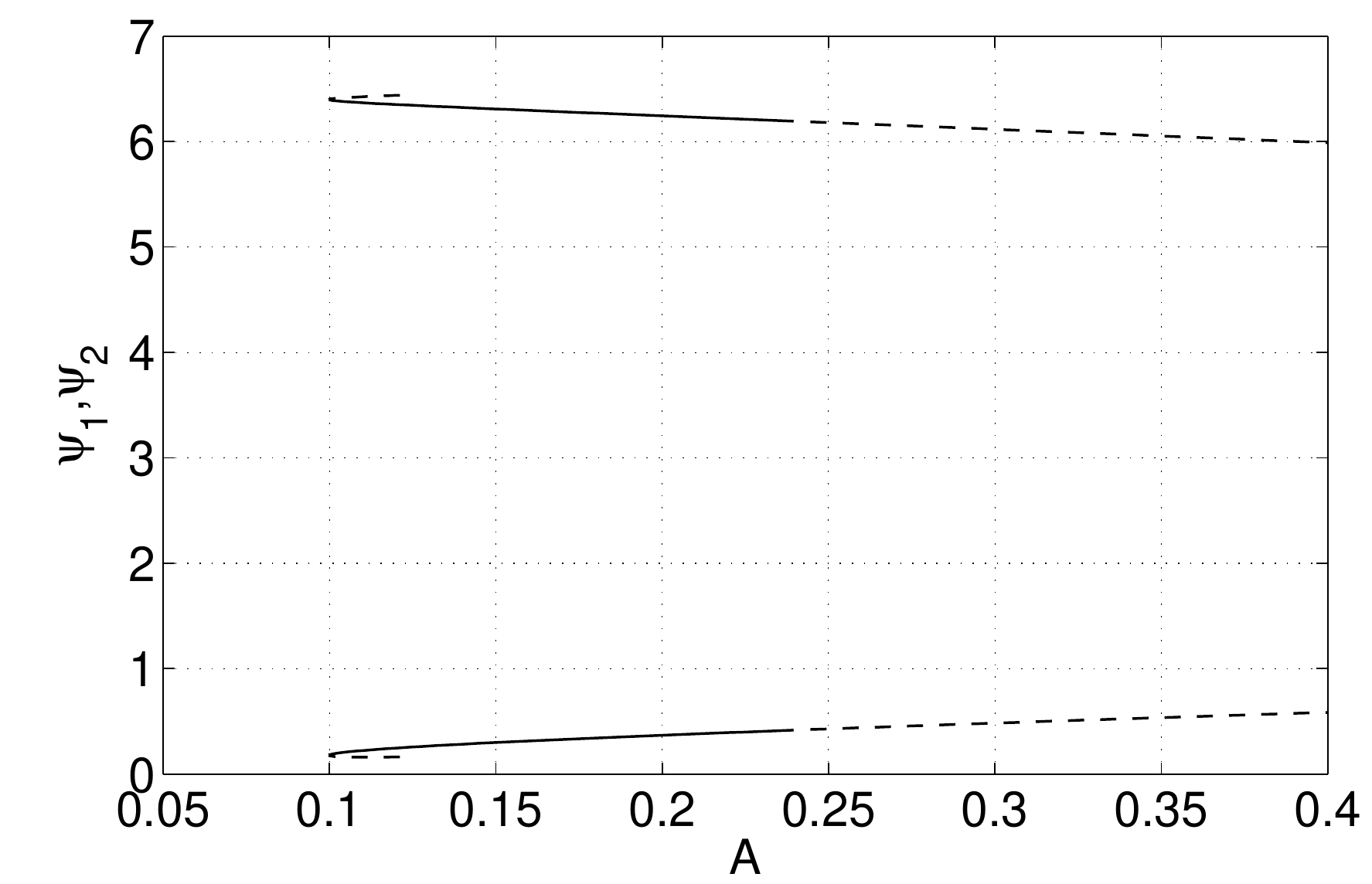}
\includegraphics[scale=0.4,clip=true,trim= 0cm 0cm 0cm 13.4cm]{fig8_follchim3-eps-converted-to.pdf}
\caption{Top: the value of $\psi_1,\psi_2$ on the Poincar{\'e} section $\psi_3=\pi$. Solid: stable, dashed: unstable.
Bottom: period of periodic orbits. (Further continuation of the higher-period branch was not possible for numerical reasons.)
Parameters: $N=3,\beta=0.1$.}
\label{fig:follchim3}
\end{figure}

Setting $N=3$ in~\eqref{sys_rot} we find stable chimera states for some parameter values. In these states, all $\psi_i$ increase monotonically in time, and thus we can reduce the dimensionality of the system by again placing a Poincar{\'e} section in the flow at, say, $\psi_3 \bmod 2\pi =\pi$. The chimera state corresponds to a fixed point of the dynamics mapping this section to itself under the dynamics of~\eqref{sys_rot}. Following such a fixed point as $A$ is varied we obtain Fig.~\ref{fig:follchim3}. As in the $N=4$ case, a stable and unstable chimera are created in a sadle-node bifurcation as $A$ is increased from zero, then the stable one loses stability through a Hopf bifurcation.

\begin{figure}[t]
\includegraphics[scale=0.4]{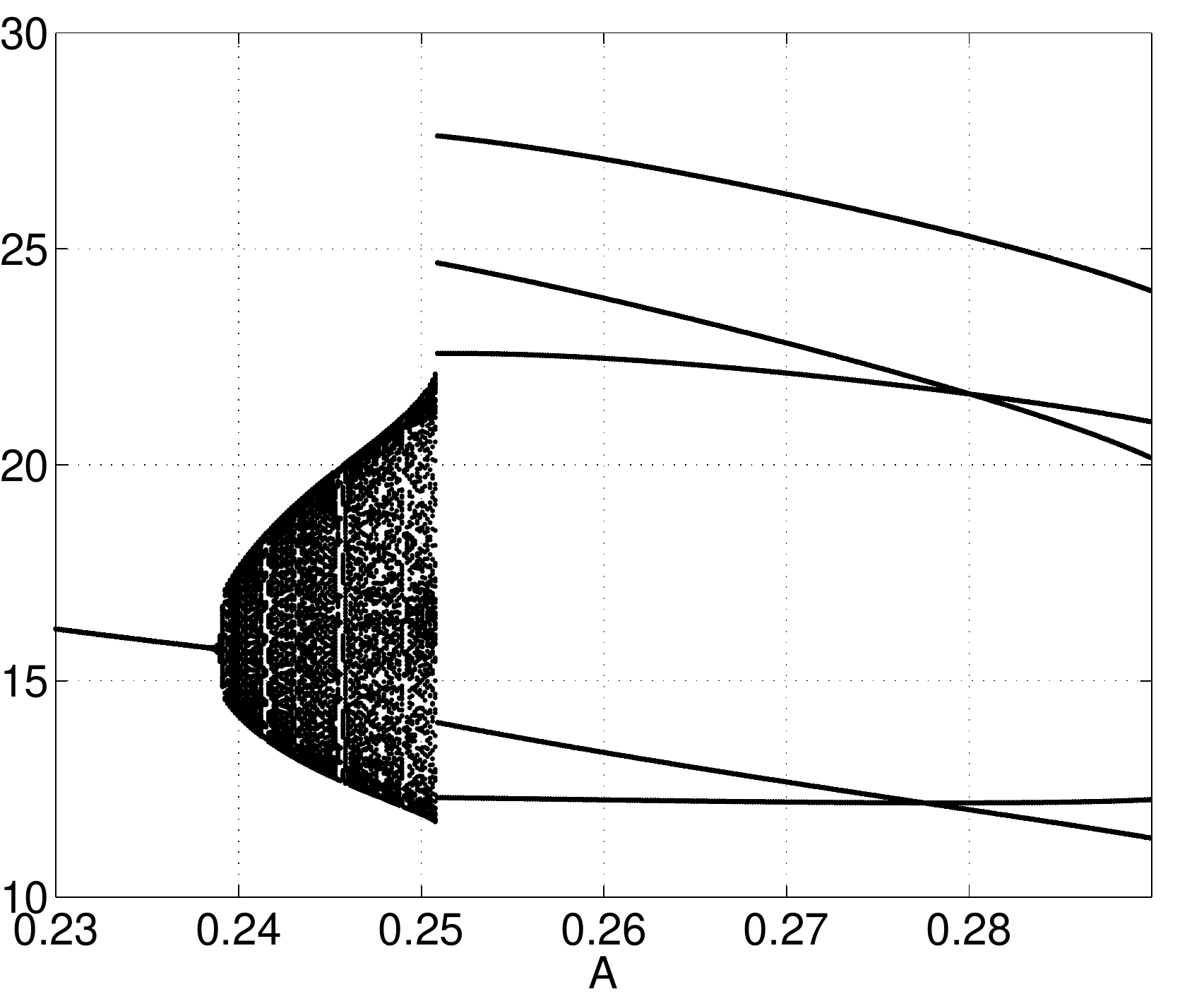}
\caption{Vertical axis: times between successive  crossings of the Poincar{\'e} section $\psi_3=\pi$ for a family of attractors corresponding to chimeras. 
Parameters: $N=3,\beta=0.1$.}
\label{fig:perA}
\end{figure}

\begin{figure}[t!]
\includegraphics[scale=0.4]{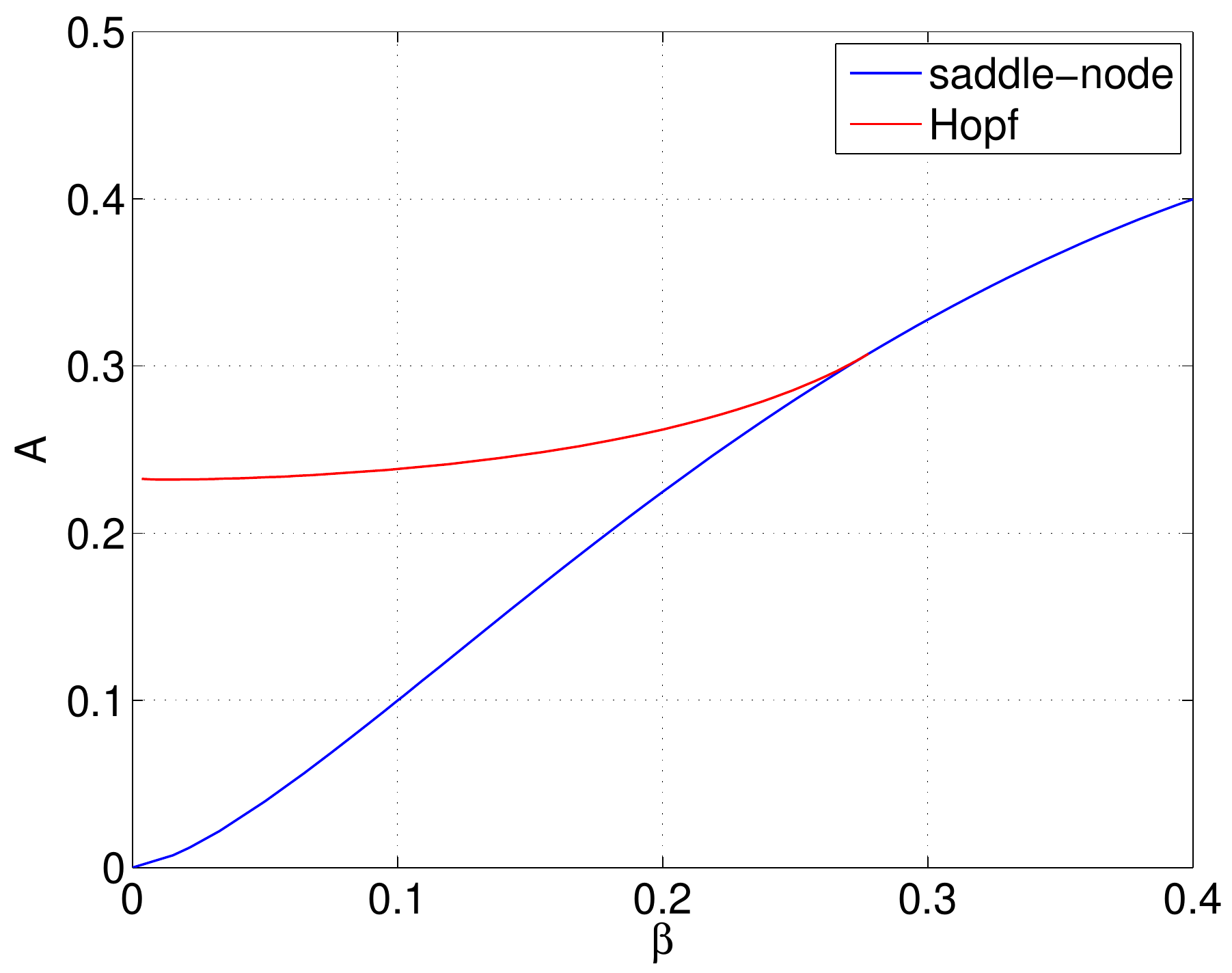}
\caption{Saddle-node and Hopf bifurcation curves for $N=3$ (see Fig.~\ref{fig:follchim3}).}
\label{fig:AbetaN=3}
\end{figure}

Increasing $A$ beyond the Hopf bifurcation and recording the times between successive crossings of the Poincar{\'e} section $\psi_3=\pi$ we obtain Fig.~\ref{fig:perA}, which suggests that the Hopf bifurcation is supercritical. However, the bifurcation associated with the appearance of the apparent period-5 orbit and its fate as $A$ is increased further are unclear. Following the saddle-node and Hopf bifurcations as $A$ and $\beta$ are varied we obtain Fig.~\ref{fig:AbetaN=3}. These curves of local bifurcations qualitatively form the same arrangement as in Figs.~\ref{fig:danny08} and~\ref{fig:wsA4}, for $N=\infty$ and $N=4$, respectively. As in the $N=4$ case, numerical investigation of the full system~\eqref{eq:dthdt}-\eqref{eq:dphidt} for $N=3$ suggests that the chimera states marked as stable in Fig.~\ref{fig:follchim3} do correspond to stable chimeras in~\eqref{eq:dthdt}-\eqref{eq:dphidt}.

\section{Dynamics for two groups of $N=2$ phase oscillators}

For general identical phase oscillator networks with fewer than four identical phase oscillators, chimera states in our sense cannot occur; see for example \cite{Ashwin2015}. 
The simplest case of our system (two groups of $N=2$ clusters) in which a chimera state can occur is therefore a network with 4 oscillators. For convenience, we rescale time by a factor of $(1+A)/2$ and consider the specific case of Eq.~\eqref{sys_rot} with $N=2$:
\begin{equation}\label{sys2}
\begin{aligned}
  \frac{d\psi_1}{d t}&=\frac{1}{2}\left[\cos(\beta)-\cos(\psi_1-\psi_2-\beta)\right]\\
  &\hspace{-5mm}+\eta(A)\left[\frac{1}{2}\{\cos(\psi_1+\beta)+\cos(\psi_2+\beta)\}-\cos(\psi_1-\beta)\right]\\
  \frac{d\psi_2}{d t}&=\frac{1}{2}\left[\cos(\beta)-\cos(\psi_2-\psi_1-\beta)\right]\\
  &\hspace{-5mm}+\eta(A)\left[\frac{1}{2}\{\cos(\psi_2+\beta)+\cos(\psi_1+\beta)\}-\cos(\psi_2-\beta)\right]
\end{aligned}
\end{equation}
where $\eta(A)=(1-A)/(1+A)$.
 
Note that Eq.~\eqref{sys2} is invariant under symmetries $\psi_1\rightarrow \psi_1+2\pi n$ and $\psi_2\rightarrow \psi_2+2\pi n$ where $n$ is an integer. It is also invariant under the symmetry $(\psi_1,\psi_2)\rightarrow (\psi_2,\psi_1)$. Any solution will either be preserved by these symmetries, or mapped to a new solution by the symmetry. Hence, any solution above the line $\psi_1=\psi_2$ will have a symmetric solution below the line as well as translates shifted by $2\pi n$ in either direction. 

We use sum/difference coordinates
\begin{equation}
\label{eq:defsumdiff}
\sigma=\frac{\psi_1+\psi_2}{2},~~\delta= \frac{\psi_1-\psi_2}{2}
\end{equation}
to write \eqref{sys2} in the more compact form:
\begin{equation}
\label{eq:sumdiff}
\begin{aligned}
\frac{d\sigma}{dt} & =  \sin^2 \delta \cos \beta  - 2 \eta(A) \sin \sigma \cos \delta \sin \beta \\
\frac{d\delta}{dt} & =  -\sin{\delta}[\sin{\beta}\cos{\delta}+\eta(A)\sin{(\beta-\sigma)}] 
\end{aligned}
\end{equation}

\subsection{Invariant structures in phase space for $N=2$}

For $N=2$ the system has an integral of the motion when $\beta=0$,
\begin{equation}
L(\sigma,\delta) := \cos \delta- \eta(A) \cos \sigma. 
\label{eq:Linv}
\end{equation}
This can be seen by computing
\begin{align}
  \frac{dL}{dt} &= \sin \beta \bigl[\sin^2\delta \left\{\cos \delta+\eta(A) \cos \sigma\right\} \nonumber\\ 
  &-2 \eta(A)^2 \cos \delta \sin^2\sigma \bigr]
\label{eq:Linvdot}
\end{align}
and observing that for $\beta=0$ we have $\frac{dL}{dt}=0$
(in the case $A=0$ this reduces to the Strogatz-Watanabe constant of the motion \cite{Watanabe1994} $L(\psi_1,\psi_2)= 2\sin \frac{\psi_1}{2}\sin \frac{\psi_2}{2}$).

\begin{figure}[ht!]
\center
\includegraphics[width=0.8\columnwidth]{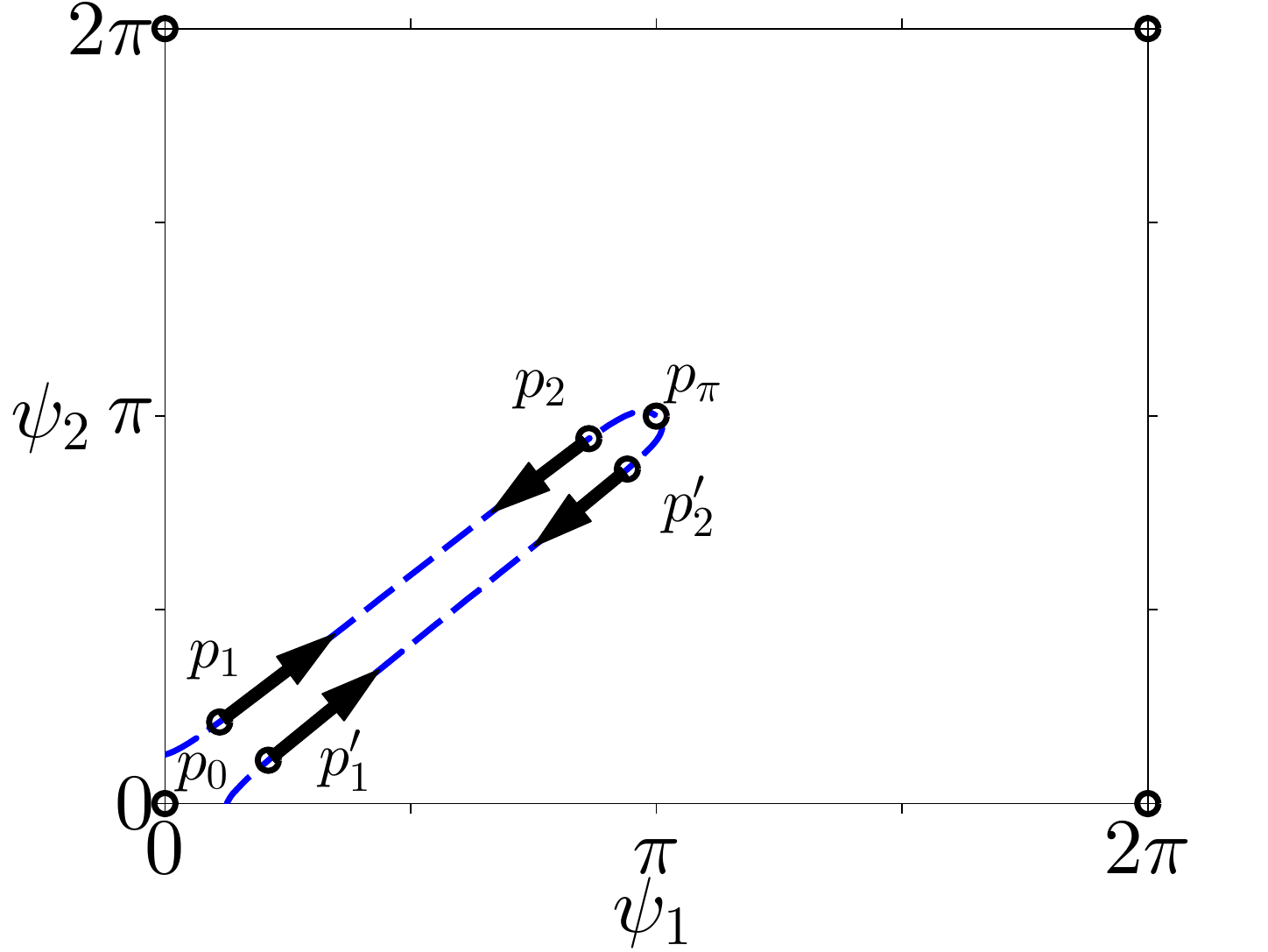}	
	\caption{Schematic of equilibria for Eq.~\eqref{sys2}. As $A$ increases, equilibria $p_1$, $p_2$, $p_1^\prime$, and $p_2^\prime$ move along the curve (blue dashed) in the direction of the displayed arrows. Ultimately, both pairs of equilibria collide and cease to exist in a saddle node bifurcation.}
	\label{fig:schematic}
\end{figure}

In addition, the symmetry of interchanging $(\psi_1,\psi_2)$ means that the line $\delta=0$ (which corresponds to $\psi_1=\psi_2=\sigma$) is invariant for all and $A$, $\beta$. Along this manifold, the flow is governed by 
\begin{equation}
\frac{d\sigma}{dt}=-2\eta(A)\sin{\sigma}\sin{\beta}.
\end{equation}
Two fixed points lie on this manifold: $p_0$, for which $\psi_1=\psi_2=0$, representing the fully synchronized solution, and $p_{\pi}$, for which $\psi_1=\psi_2=\pi$, representing a solution in which oscillators within each group are synchronized but the two groups are $\pi$ out of phase (antiphase) with each other (see Fig.~\ref{fig:schematic}). For $0\leq \beta \leq \pi/2$, the domain where chimera states are typically of interest, the fully synchronized solution $p_0$ is stable while the out of phase solution, $p_\pi$, is a saddle.  

Linearizing \eqref{eq:sumdiff} about the manifold $\delta=0$ gives
\begin{equation}
\frac{d\delta}{dt}=-\left[\eta(A)\sin{(\beta-\sigma)}+\sin{\beta}\right]\delta+ O(\delta^2),
\end{equation}
and thus this manifold may be repelling for some $\sigma$ and attracting for others, depending on the values of $A$ and $\beta$. 
For parameters of interest, there are also two pairs of fixed points with $\psi_1 \neq \psi_2$, one pair, $p_1$ and $p_1^{\prime}$, near (0,0) and another pair, $p_2$ and $p_2^{\prime}$, near $(\pi,\pi)$ (see Fig.~\ref{fig:schematic}, Appendix \ref{App:FP}). $p_1$ and $p_1^{\prime}$ are mapped to one another by the operation $(\psi_1,\psi_2)\mapsto(\psi_2,\psi_1)$, as are $p_2$ and $p_2^{\prime}$. For concreteness, we say that $p_1$ and $p_2$ are above the diagonal, while $p_1^{\prime}$ and $p_2^{\prime}$ are below. All of these are unstable. 
Analysis of the Jacobian at these fixed points reveals that both $p_1$ and $p_2$ are saddles when $A$ is small. As $A$ increases, $p_2$ becomes an unstable spiral node. Eventually, at a critical value of $A$ (see Appendix \ref{App:SN}) $p_1$ and $p_2$ collide in a saddle node bifurcation. 

\begin{figure}[hb!]
\includegraphics[width=0.7\columnwidth]{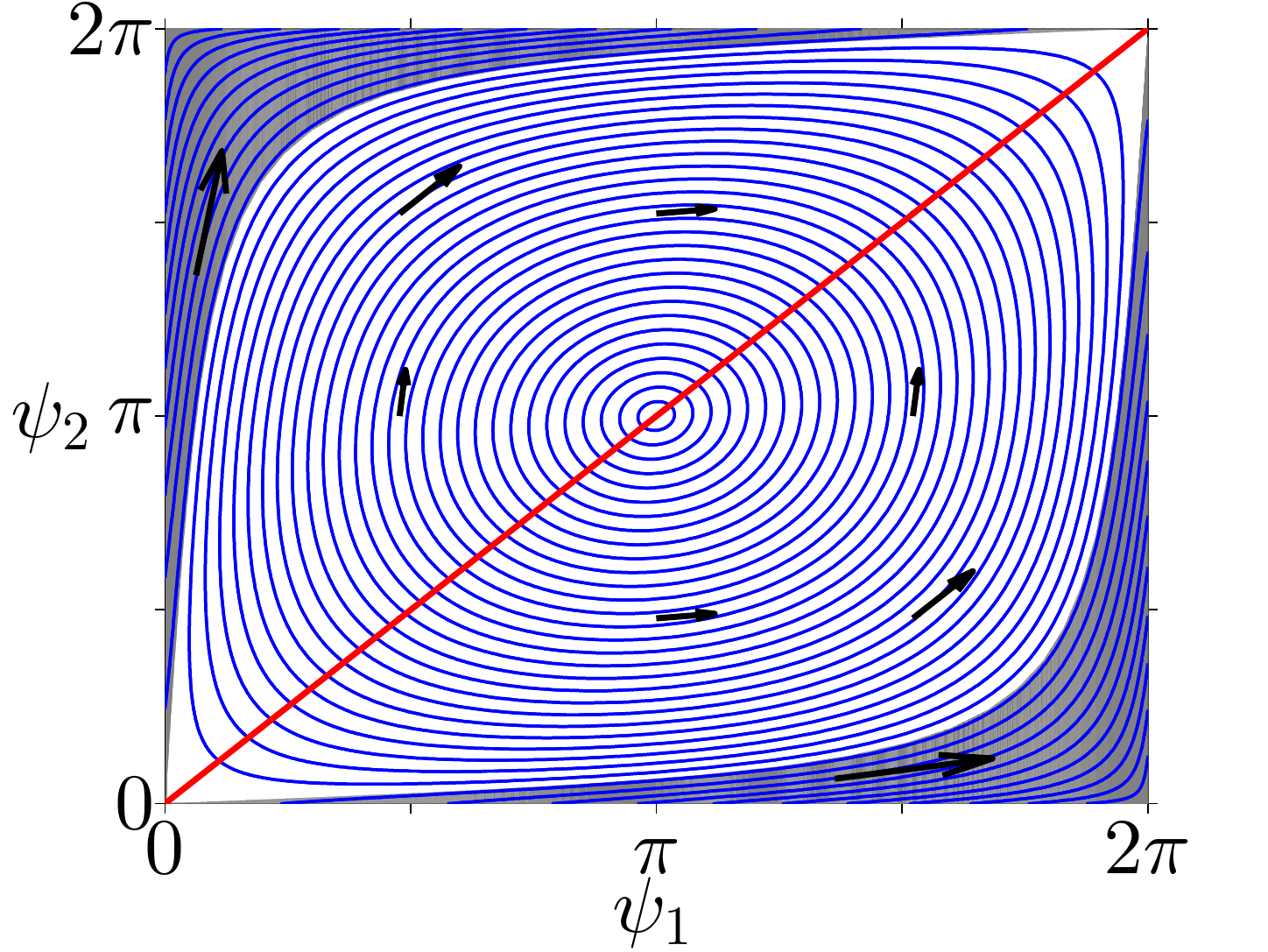}
\\	
\includegraphics[width=0.7\columnwidth]{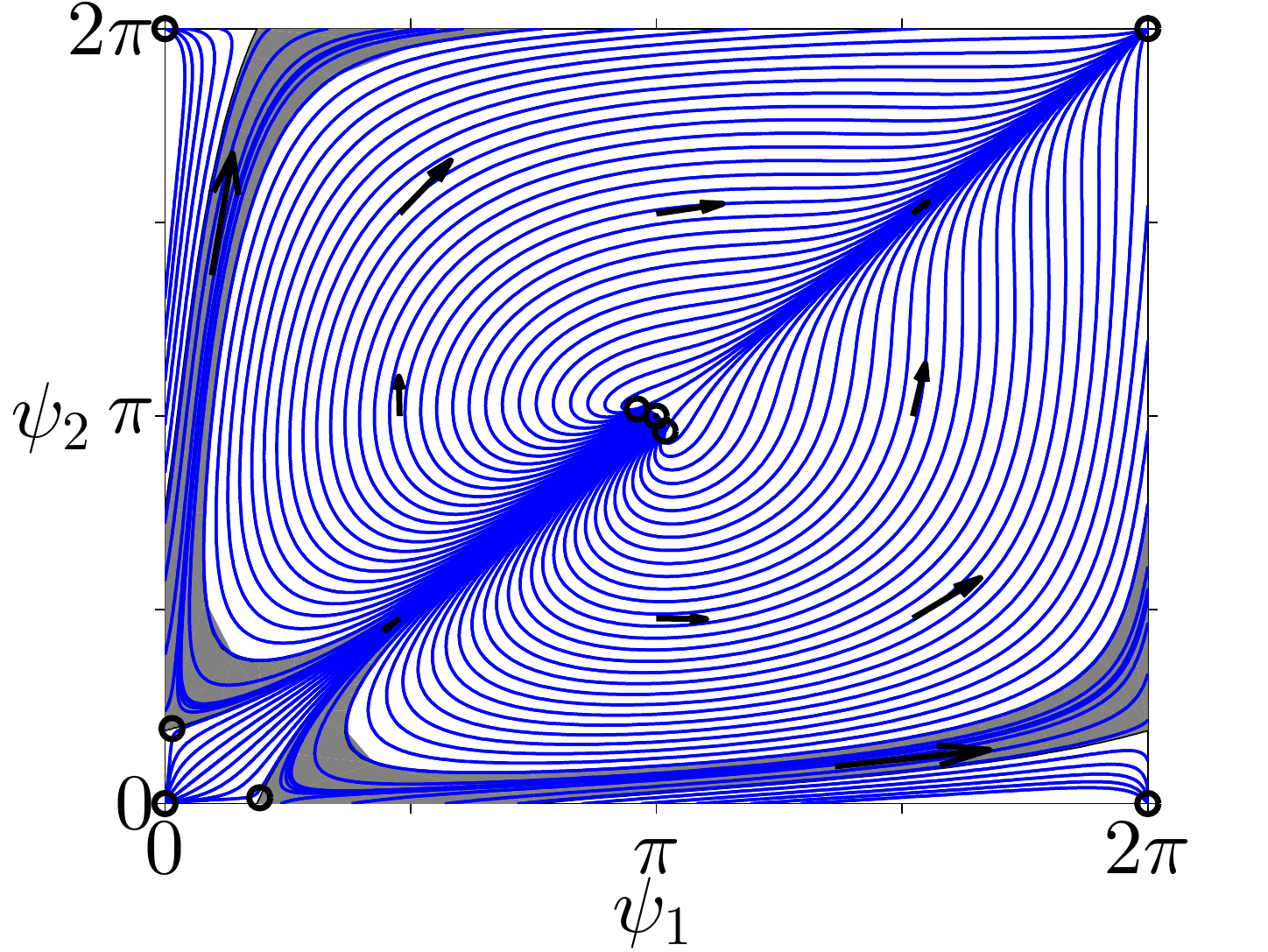}
	\caption{Trajectories for Eq.~\eqref{sys2} with $A=0.1$ and varying $\beta$. Top: For $\beta=0$, all trajectories remain on level curves of the invariant $L$ (\ref{eq:Linv}). The gray shaded regions indicate neutrally stable chimera periodic orbits while the red line is a line of neutrally stable fixed points. Bottom: (b) For $\beta=0.15$ fixed points are indicated by black circles. Trajectories within the gray band are stable chimeras.}	\label{fig:trajectories}
\end{figure}

In order to explore the solutions of Eqns.~\eqref{sys2}, we integrate the equations for a variety of initial conditions satisfying $0\leq\psi_1,\psi_2\leq2\pi$; see Fig.~\ref{fig:trajectories}. For fixed $\beta$ we observe two different types of long-term attracting behaviors. When $\beta=0$, all trajectories remain on level curves of $L$ that either connect pairs of neutrally stable fixed points or are infinitely long periodic orbits.  For $0<\beta\ll 1$, most initial conditions evolve toward the fully synchronized state ($p_0$).  However, there is a set of initial conditions that give rise to a stable chimera solution where $\psi_1$ and $\psi_2$ increase indefinitely at an asymptotically linear rate, and for which $\psi_1(t)\neq\psi_2(t)$. 
In order to understand this behavior better we now examine the $(\psi_1,\psi_2)$ phase plane as $A$ is varied.

\vspace{-0.7cm}


\begin{figure*}[th]
\includegraphics[width=\textwidth,clip=true,trim=0cm 0cm 1cm 0cm]
{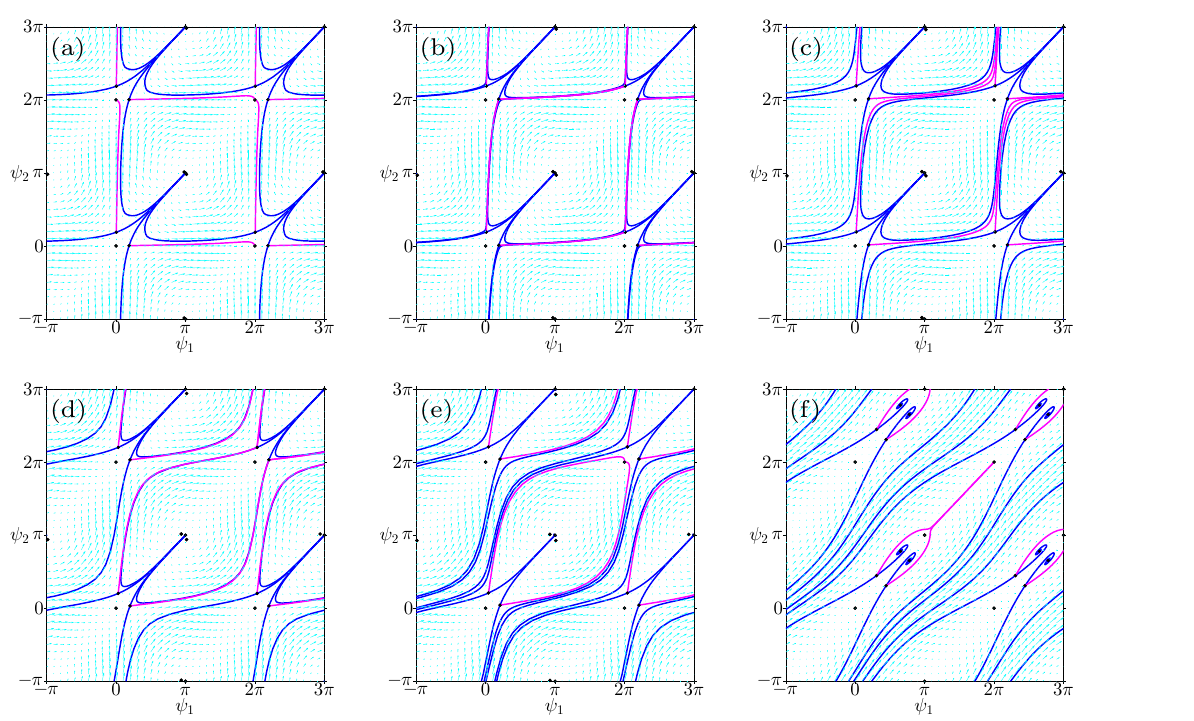}
\caption{Phase plots for Eq.~\eqref{sys2}. Each panel contains a phase plot for $\beta=0.15$ and for various values of $A$ (a) $A=0.03$, (b) $A=0.0662$, (c) $A=0.1$, (d) $A=0.194$, (e) $A=0.25$, and (f) A=0.71 Cyan arrows indicate the direction of flow in the plane.  Fixed points are marked with black circles.  Curves indicate selected stable (blue) and unstable (magenta) manifolds of the saddle points.} 
\label{fig:phaseplane_two_bif}
\end{figure*}

\subsection{Phase plane analysis for $N=2$}

\begin{figure}[ht]
\includegraphics[scale=0.275,clip=true,trim=0cm 0cm 3.15cm 0cm]
{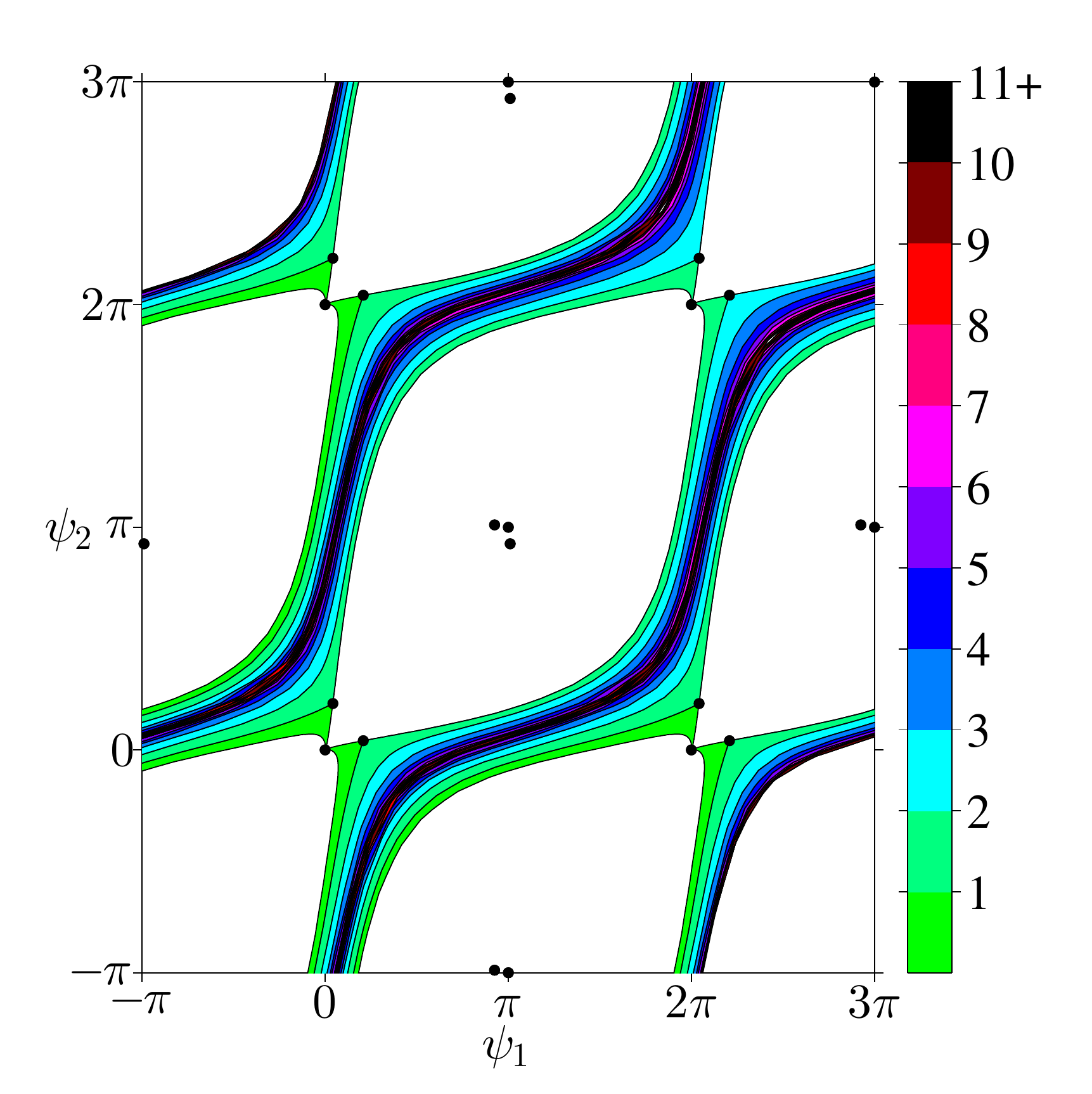}\hspace{0.1cm}
\includegraphics[scale=0.275,clip=true,trim=1.6cm 0cm 0cm 0cm]
{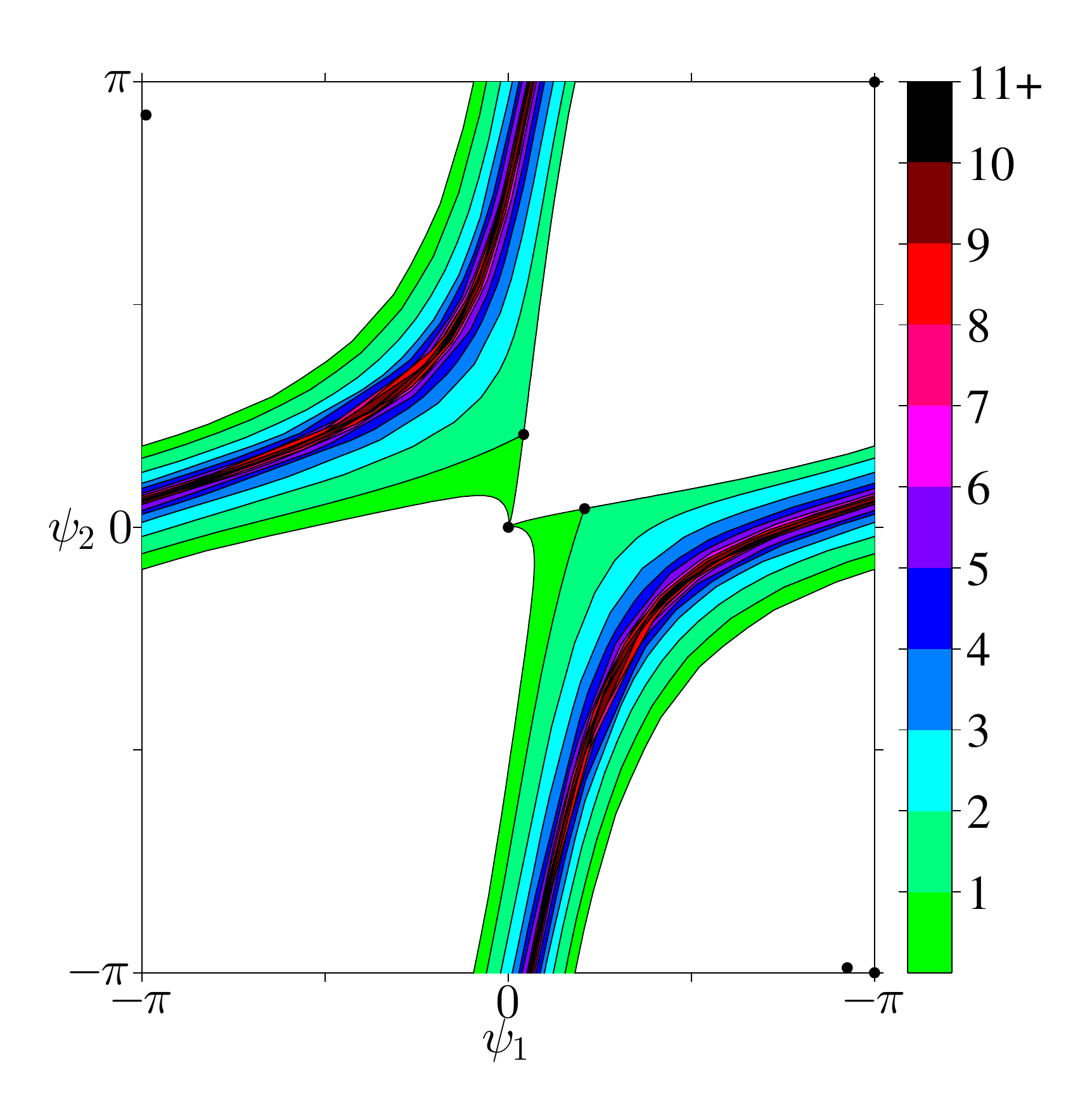}
\caption{Foliation of the phase plane. Left and right panels depict the phase plane for $A=0.25$ and $\beta=0.15$. The white region exhibits no oscillation. The colored regions contain chimera states with finite lifetimes. The number of cycles before a fixed point is reached is indicated by the color.}
\label{fig:basins}
\end{figure}

Figure \ref{fig:phaseplane_two_bif} displays the equilibrium points and some of the relevant manifolds of $p_1$ and $p_1^{\prime}$, and the direction of flow for fixed $\beta=0.15$ and six different values of $A$. This figure reveals a variety of different scenarios. When $A$ is small, all trajectories converge to $p_0$ (the fully synchronous state). As $A$ increases, a global bifurcation occurs (see Fig.~\ref{fig:phaseplane_two_bif}(b)). The nearly-vertical unstable manifold of $p_1$ (magenta) merges with part of the stable manifold of $p_1'$ (blue), where by ``$p_1'$'' we mean, in this case, $p_1'$ shifted vertically by $2\pi$. This results in the creation of a periodic orbit with an infinite period.  Beyond this bifurcation, a narrow ``channel'' bounded by the unstable manifolds of $p_1$ and $p_1'$ forms. Initial conditions within this channel (including those along the magenta unstable manifolds) cannot approach any fixed point without crossing one of these stable manifolds (see Fig.~\ref{fig:phaseplane_two_bif}(c)),  therefore, the lifetimes of these trajectories are infinitely long. Over time, they approach a periodic orbit that represents a chimera state.  As $A$ continues to increase, eventually the nearly-vertical unstable manifold of $p_1$ merges with a stable manifold of the image of $p_1$ under the action $(\psi_1,\psi_2)\mapsto(\psi_1+2\pi,\psi_2+2\pi)$ (see Fig.~\ref{fig:phaseplane_two_bif}(d)). This causes a second global bifurcation in which the two manifolds exchange orientations (see Fig.~\ref{fig:phaseplane_two_bif}(e)).  Here the stable manifolds of $p_1$ and $p_1'$  can be traced backwards indefinitely. Thus there are arbitrarily long transients in the system (which we refer to as `backwards chimera states'), but all trajectories eventually approach $p_0$. The lifetimes of these transients are displayed in Fig.~\ref{fig:basins}.  Eventually, these ``channels'' completely vanish when $p_1$ and $p_2$ disappear as $p_1$ and $p_2$ approach each other (see Fig.~\ref{fig:phaseplane_two_bif}(f)) and undergo a saddle node bifurcation.

\subsection{Bifurcations for $N=2$}

\begin{figure}[ht]
\includegraphics[scale=0.4,clip=true,trim= 0cm 13.6cm 0cm 0cm]{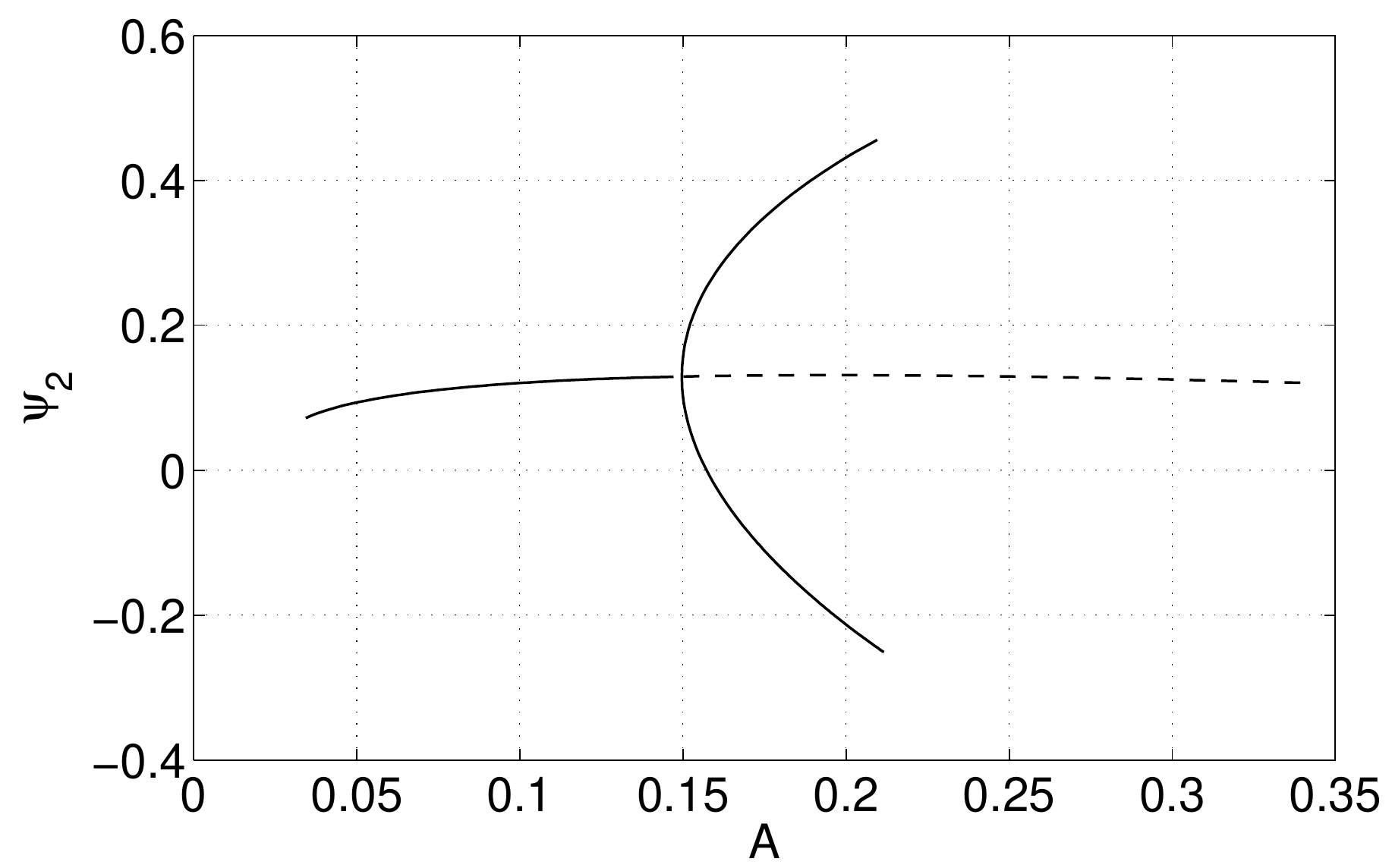}
\includegraphics[scale=0.4,clip=true,trim= 0cm 0cm 0cm 13.4cm]{fig15_follchim-eps-converted-to}
\caption{Top: the value of $\psi_2$ on the Poincar{\'e} section $\psi_1=\pi$. Solid: stable, dashed: unstable.
Bottom: period of periodic orbits. (Continuation to higher periods was not possible for numerical reasons.)
Parameters: $\beta=0.1$.}
\label{fig:follchim}
\end{figure}

The bifurcations described above appear for a range of $\beta$ values.  As is the case for $N=3$, the parameter values at which these bifurcations occur can be computed numerically by placing a Poincar{\'e} section in the flow, say at $\psi_1\bmod 2\pi=\pi$. Following fixed points of the map from this section to itself we obtain the results in Fig.~\ref{fig:follchim}. We see that the bifurcations observed when $N=2$ differ from those found above for $N=3,4$ and in Ref.~\cite{Abrams2008} for $N\rightarrow\infty$. Stable  periodic orbits in the flow first appear at the global bifurcation involving the unstable manifold of $p_1$, as shown in Fig.~\ref{fig:phaseplane_two_bif}(b).  As $A$ increases, a supercritical pitchfork bifurcation of this orbit occurs and the stable periodic orbit splits into a pair of stable orbits and an unstable orbit. Eventually, both of the stable chimera states disappear when the second global bifuration occurs, as shown in Fig.~\ref{fig:phaseplane_two_bif}(d).  The unstable periodic orbit persists for larger $A$.

\begin{figure}[th]
\includegraphics[width=0.9\columnwidth]{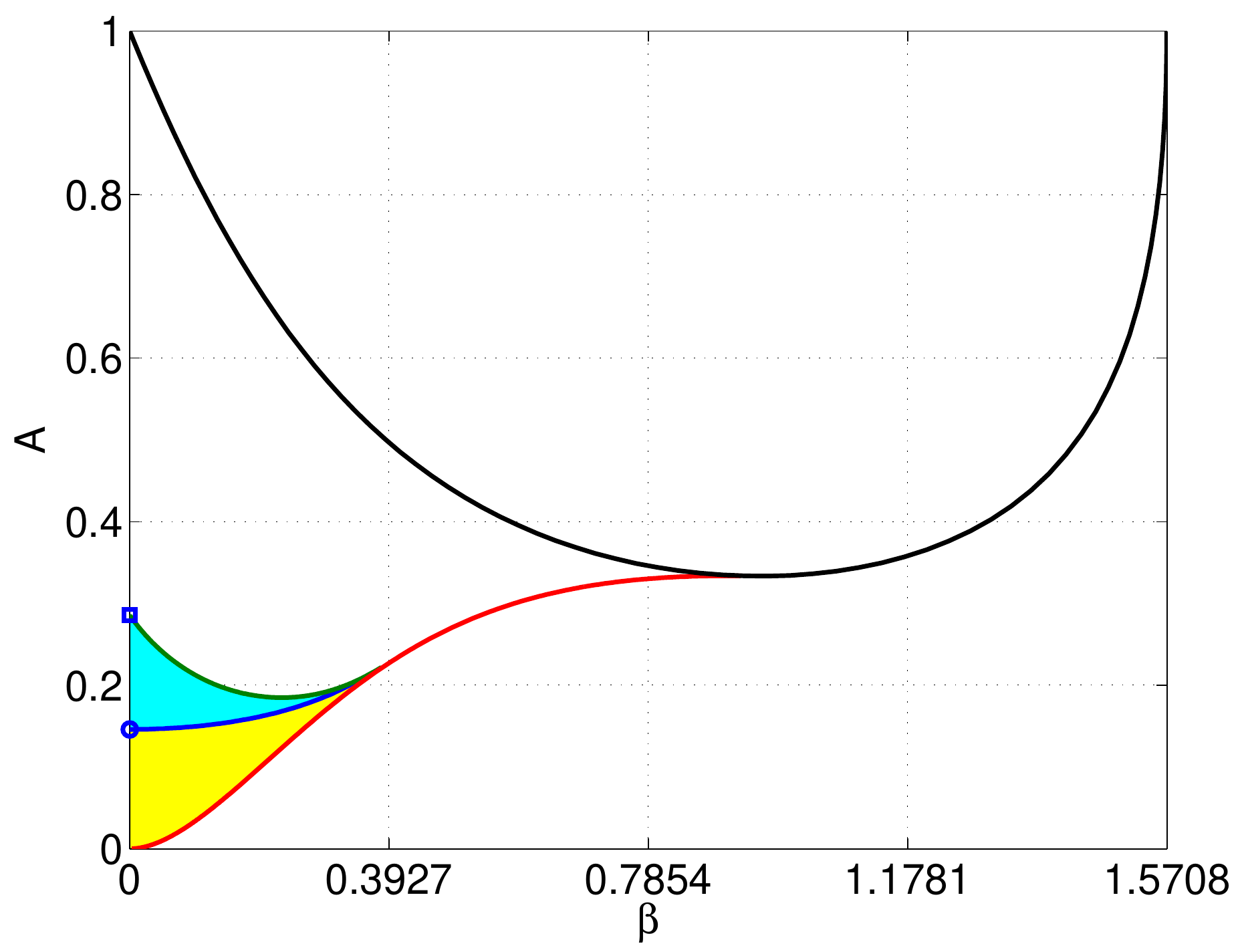}
\caption{Bifurcations for the $N=2$ case. Black: saddle-node of fixed points; blue: pitchfork of chimeras;
red: heteroclinic connection between $p_1$ and $p_1'$ (with one variable increased by $2\pi$); green:
homoclinic connection between $p_1$ and $p_1$ (with both variables increased by $2\pi$). Stable chimeras exist in the shaded regions bounded by the green and red curves, and the line $\beta=0$. The blue circle ($A_0\approx 0.145898$) and blue square ($A_1\approx0.285714$) indicate analytical predictions for the bifurcation points.}
\label{fig:Abeta}
\end{figure}

Following the bifurcations shown in Fig.~\ref{fig:follchim} we obtain Fig.~\ref{fig:Abeta}. Although the bifurcations involved in creating and destroying chimera states are different than for the $N>2$ cases, surprisingly, two types of stable chimeras still exist in a ``wedge'' in parameter space with one corner at the origin in $(A,\beta)$ space. Although only the results of numerical calculations have been shown so far, in Appendix \ref{sec:appendixN2} we present analytical calculations of the left boundaries of the blue and green curves in Fig.~\ref{fig:Abeta}, $(A_0,0)$ and $(A_1,0)$, respectively.

\section{Discussion/conclusion}

Chimera states have been observed in a variety of networks~\cite{Kuramoto2002, Abrams2004, Shima2004, Abrams2006,Martens2010_1, Martens2010_2, Omelchenko2012, Zhu2012, Panaggio2013,Panaggio2014,Laing2009_2, Laing2012, Yao2013,Hagerstrom2012,Tinsley2012,Nkomo2013} but perhaps the simplest network in which they are seen is one consisting of two groups of $N$ oscillators each, with all-to-all coupling within and between groups~\cite{Abrams2008,Laing2009_1,Laing2012_2,Martens2013,Pikovsky2008}. Most analysis of chimera states has been in the continuum limit --- where $N$ is taken to be infinite --- but here we have investigated the opposite limit, seeing how few oscillators were needed in a network in order to observe what could be referred to as a chimera. Surprisingly, we found that chimera states exist and are stable even for $N=2$, and that the bifurcation scenarios for $N>2$ are qualitatively the same as for $N=\infty$.

For $N>3$ we used the Watanabe/Strogatz transformation and the prior calculations of Pikovsky and Rosenblum~\cite{Pikovsky2008} to derive the three ODEs~\eqref{eq:drhodt}-\eqref{eq:dpsidt} which describe the asynchronous group of oscillators. For $N=4$ we found a bifurcation scenario which was qualitatively the same as that for the $N=\infty$ case. 
For $N=3$ we studied the three equations~\eqref{sys_rot} for the phases differences between the two groups and also found a bifurcation scenario which was qualitatively the same as that for the $N=\infty$ case. The $N=2$ case is described by the two ODEs~\eqref{sys2} and the bifurcation scenario corresponding to this system is quite different from all other cases, as it involves global bifurcations and a pitchfork bifurcation. 

Note that the chimeras discussed here are ``weak chimeras'' in the sense of~\cite{Ashwin2015}. In that paper, for the case $N=2$ and a set of phase oscillators that reduces to the problem considered here for a special choice of parameters, attracting weak chimeras are found even in the limit $\nu\rightarrow 0$ for fixed $\mu$. 

For chimera states to be observable they must be robust with respect to hetereogeneity. Previous investigations have shown that they are robust with respect to non-identical frequencies~\cite{Laing2009_1,Laing2009_2} (in the continuum limit) and removal of connections within a large but finite network~\cite{Laing2012_2}. The derivation of many of the equations in this paper has relied upon the oscillators having identical intrinsic frequencies, but we have verified numerically that the stable chimera states found here do persist when the intrinsic frequencies are randomly chosen from distributions with sufficiently small variances, for all $N\geq 2$ (results not shown). 

Another question that arises is whether these results can be extended to networks of oscillators which are not described by just a phase variable. Clearly chimera states are possible in these systems, as some of the first observations of chimera states were in networks of such oscillators~\cite{Kuramoto2002,Shima2004} (also see~\cite{Laing2010,Kawamura2007,Sakaguchi2006,Kuramoto2003_2}). However, all of these studies considered large (or infinite) networks, and there is an open question as to whether the chimera states found here in small networks of phase oscillators exist in small networks of more general oscillators.



\appendix


%
%

%

\section{Stability of the synchronized group}\label{App:Stab}
In order to show that the manifold in which one group is synchronized is stable, we begin by considering considering equations \eqref{eq:dthdt}-\eqref{eq:dphidt}. Moving into a rotating frame of reference, we let $\Phi_i=\phi_i-\phi$ and $\psi_i=\theta_i-\phi$ where 
\begin{align*}
\frac{d\phi}{dt}=\omega-\frac{1+A}{2}\cos{\beta}-\frac{1-A}{2N}\sum_{j=1}^N\cos(\psi_j+\beta).
\end{align*}
After rescaling time by a factor of $(1+A)/2$ and defining $\eta(A)=(1-A)/(1+A)$, we obtain 
\begin{align}
  \frac{d\psi_i}{d t} &= \left[\cos\beta-\frac{1}{N}\sum_{j=1}^N\cos(\psi_i-\psi_j-\beta)\right] \nonumber\\
  &+\frac{\eta(A)}{N} \left[\sum_{j=1}^N\cos(\psi_j+\beta) - \sum_{j=1}^N\cos(\psi_i-\Phi_j-\beta)\right], \label{eq:dthdt2} \\
  \frac{d\Phi_i}{d t} &= \left[\cos\beta-\frac{1}{N}\sum_{j=1}^N\cos(\Phi_i-\Phi_j-\beta)\right] \nonumber\\
  &+\frac{\eta(A)}{N}\left[\sum_{j=1}^N\cos(\psi_j+\beta)-\sum_{j=1}^N\cos(\Phi_i-\psi_j-\beta)\right]. \label{eq:dphidt2}
\end{align}
Note that $\Phi_i=0$ is an invariant manifold in equation \eqref{eq:dphidt2}. We now derive conditions on $\psi_j,A,\beta$ such that this state is stable.  To achieve this, we treat $\psi_j$ as external forcing functions and compute the Jacobian of equation \eqref{eq:dphidt2}. It is straightforward to show that along the synchronized manifold ($\Phi_i=0$) the Jacobian satisfies
\begin{align*} J_{ii}&=-\left(\frac{N-1}{N}\right)\sin(\beta)-\left(\frac{\eta(A)}{N}\right)\sum_{j=1}^N\sin(\psi_j+\beta),\\
J_{ij}&=	\displaystyle\frac{1}{N}\sin(\beta), \quad j\neq i 
\end{align*}
For compactness, we define $Z=\frac{1}{N}\sum_{j=1}^N\sin(\psi_j+\beta)$.
In matrix form, the Jacobian can be expressed as 
\[J=-\frac{1}{N}\sin(\beta)L_N-\eta(A)ZI_N\]
where $I_N$ is the $N\times N$ identity matrix and \[L_N=\left(\begin{array}{ccccc}
N-1 & -1  & \ldots &  -1 \\
-1 & N-1  & \ddots & \vdots  \\
\vdots & \ddots & \ddots  &-1 \\
-1 & \ldots &-1 & N-1  
\end{array}\right)\] is the Laplacian of a complete graph with $N$ vertices. Making use of the fact that $L_N$ has a single zero eigenvalue and a repeated eigenvalue $N$ (with multiplicity $N-1$), it is straightforward to show that the eigenvalues of $J$ are:
\[\lambda_1=-\eta(A)Z,\quad\lambda_2=-\sin(\beta)-\eta(A)Z\]
The eigenvector corresponding to $\lambda_1$ is $\mathbf{1}$ representing a uniform phase shift along the synchronized manifold.  Thus the stability transverse to the manifold is determined by $\lambda_2$. Rearranging, we observe that $\lambda_2<0$ is equivalent to 
\[Z>-\frac{\sin\beta}{\eta(A)}.\]
$Z$ depends on the phases of the oscillators in the desynchronized group and its value cannot be determined a priori. However, it is clearly bounded by $-1\leq Z\leq 1$. So the synchronized manifold is always stable for $\eta(A) < \sin \beta$, or 
\[
  A>\frac{1-\sin\beta}{1+\sin\beta}.
\]
In practice, however $\left|Z\right|$ remains close to 0 in a chimera state, so the synchronous group is stable to a variety of perturbations provided $\frac{\sin\beta}{\eta(A)}$ is not too small.
\section{Asymptotic expressions for fixed points $p_1$ and $p_2$} \label{App:FP}

When $A\ll 1$, the coordinates of fixed points $p_1$ and $p_2$ can be approximated as follows:

Fixed point $p_1$ scales like  
\begin{align*}
\psi_1&\sim\sin(\beta)\cos(\beta)\left[3A+\frac{5}{4}\left(3\sin^2(\beta)+1\right)A^2  \right.\\
&\quad\left.-\frac{1}{8}\left(3\sin^4(\beta)-38\sin^2(\beta)-21\right)A^3\right]\\
\psi_2&\sim \psi_2^0+\sin(\beta)\cos(\beta)\left[A+\frac{1}{4}\left(\sin^2(\beta)+11\right)A^2 \right.\\
&\quad\left.-\frac{1}{24}\left(55\sin^4(\beta)-142\sin^2(\beta)-33\right)A^3\right]\\
\end{align*}
where $\psi_2^0$ satisfies the equation
\begin{equation*}
\tan(\beta)=\frac{1-\cos(\psi_2^0)}{2\sin(\psi_2^0)}.
\end{equation*}

Similarly, fixed point $p_2$ scales like 
\begin{align*}
\psi_1&\sim\pi-2\sin(\beta)A^{\frac{1}{2}}-2\sin(\beta)\cos(\beta)A \\
&\quad-\left(\frac{1}{3}\sin(\beta)^3-\sin(\beta)\right)A^{\frac{3}{2}}\\
\psi_2&\sim\pi+2\sin(\beta)A^{\frac{1}{2}}-2\sin(\beta)\cos(\beta)A \\
&\quad+\left(\frac{1}{3}\sin(\beta)^3-\sin(\beta)\right)A^{\frac{3}{2}}.
\end{align*}
These expressions are useful for identifying initial conditions that lead to chimera states.

\section{Saddle node bifurcation curve for $N=2$}\label{App:SN}


In order compute the saddle node bifurcation curve shown in Fig.~\ref{fig:Abeta}, we now consider the Jacobian of Eq.~\eqref{sys2} near $p_1$. At the bifurcation, the determinant be must equal to 0, so in theory one can find a single equation in $A$ and $\beta$ corresponding to the saddle node curve.  Unfortunately, thus far we have been unable to find a closed form solution for the fixed points. So, instead, we look for a perturbative expression when $A\sim1$ and $\beta\ll1$.  
We assume that the solution satisfies the following where $\epsilon\ll1$ is a small parameter
\begin{align*}
A&=1-\epsilon\\
\beta&\sim\epsilon\beta_1+\epsilon^2\beta_2+\epsilon^3\beta_3+\epsilon^4\beta_4\\
\psi_i&\sim \psi_{i_0}+\epsilon\psi_{i_1}+\epsilon^2\psi_{i_2}+\epsilon^3\psi_{i_3}+\epsilon^4\psi_{i_4}.
\end{align*} 
Substituting this into the determinant of the Jacobian, expanding in $\epsilon=(1-A)$, we obtain the solution curve
\begin{equation*}
\beta\sim\frac{125}{96}-\frac{47}{16}A+\frac{11}{4}A^2-\frac{67}{48}A^3+\frac{9}{32}A^4.
\end{equation*}
which agrees with the numerical results displayed in Fig.~\ref{fig:Abeta} as $A\rightarrow 1$.

\section{Bifurcations of chimera for $N=2$ near $\beta=0$} \label{sec:appendixN2}

The solutions and bifurcation curves described above for Eq.~\eqref{sys2} can be computed analytically in the limit $0<\beta\ll1 $. We compute using the Eq.~\eqref{eq:sumdiff} in the sum/difference coordinates \eqref{eq:defsumdiff} and 
exploit the fact that $L$ defined in (\ref{eq:Linv}) is, by (\ref{eq:Linvdot}), close to an invariant for this limit.

In particular, for $\beta=0$ the level curve $L=L_0$ is invariant and we can eliminate $\delta$
\begin{equation}
\begin{aligned}
  \cos \delta &= \eta(A)\cos \sigma + L_0,~~\\
  \sin^2 \delta &= 1- (\eta(A) \cos \sigma + L_0)^2,~~
\end{aligned}
\end{equation}
where $\eta(A)=(1-A)/(1+A)$ meaning that
\begin{equation}
\frac{d}{dt} \sigma = 1- (\eta(A)\cos \sigma + L_0)^2
\label{eq:sigmaL0}
\end{equation}
and
$$
\delta = \Delta(A,\sigma,L_0):=\arccos\left(\eta(A) \cos \sigma+L_0\right).
$$
All trajectories with $|L_0|<1-\eta(A)$ will wind around $\sigma$ but not $\delta$; they are ``weak chimeras'' and are periodic orbits with a period $T(L_0)$ that can be found in integral form.

Now consider $0<\beta\ll 1$ and note that
\begin{equation}
  \frac{dL}{dt} = \beta G(\sigma,\delta) + O(\beta^2)
\end{equation}
where
\[
  G(\sigma,\delta) = \sin^2\delta \left(\cos \delta+\eta(A) \cos \sigma\right) -2\eta(A)^2 \cos \delta \sin^2\sigma.
\]
Hence the change in $L$ over one period $T(L_0)$ on $L_0$ is determined to lowest order in $\beta$ by
\begin{equation}
\Lambda(L_0):=\int_{t=0}^{T(L_0)} G(\sigma,\Delta(A,\sigma,L_0)) dt
\end{equation}
in the sense that $L(t+T(L(t)))= L(t)+\beta\Lambda(L(t))+O(\beta^2)$ for fixed $A$ and small $\beta$, for any region of $L$ such that $T(L)$ is bounded. If there is a (weak chimera) periodic orbit within $\beta$ of $L=L_0$ then $\Lambda(L_0)=0$. The stability of a (weak chimera) periodic solution $L_0$ is determined by a Floquet multiplier $1+\beta M(L_0)+O(\beta^2)$, where
$$
M(L_0):=\frac{d\Lambda}{dL}(L_0)
$$
so we have (for $0<\beta\ll 1$) linear stability if $M(L_0)<0$ and linear instability if $M(L_0)>0$. One can show that
\[\Lambda(0)= \int_{\sigma=0}^{2\pi } \frac{2\eta(A)(1-\eta(A)^2)\cos\sigma}{1- \eta(A)^2 \cos^2 \sigma}\,d\sigma = 0\]
(the integrand has symmetry $I(\sigma+\pi)=-I(\sigma)$) meaning there is a periodic solution close to the curve $L_0=0$ that is a weak chimera for $0<A$. We compute
\begin{eqnarray*}
M(0) & = & 2\pi \left(\frac{3 \sqrt{A}-1-A}{\sqrt{A}}\right)
\end{eqnarray*}
which has a unique zero for $0<A_0$ at
\begin{equation}
A_0:=\frac{7-3\sqrt{5}}{2}\approx 0.1458980337.
\end{equation}
For $0<\beta\ll 1$, if $0<A<A_0$ then $M(0)<0$ and the periodic orbit is stable while for $A>A_0$ it becomes unstable at a {\em pitchfork of (weak chimera) limit cycles}.

The {\em homoclinic connection between $p_1$ and $p_1$ (with both variables shifted by $2\pi$)} for $0<\beta\ll 1$ can similarly be found by finding a value of $A$ such that the chimera periodic orbit is at the level $L_0=2A/(1+A)$ corresponding to the heteroclinic orbit of the integrable system at $\beta=0$. On this orbit we can similarly calculate
\begin{align*}
\Lambda\left(\frac{2A}{1+A}\right) & = 2\pi  \frac{ 2A\sqrt{2+2A} - 6 A\sqrt{A}}{\sqrt{A+A^2}}
\end{align*}
and so in the limit $0<\beta\ll 1$ the bifurcation is at $\Lambda(2A/(1+A))=0$, meaning that
\begin{equation}
A_1=\frac{2}{7} \approx 0.285714.
\end{equation}
The {\em{heteroclinic connection between $p_1$ and $p_1^\prime$ (with one variable increased by $2\pi$)}} is at $A=0$ in the limit $0<\beta\ll 1$, which is where the $L_0=0$ chimera solution has unbounded period. The values of $A_0$ and $A_1$ are shown in Fig.~\ref{fig:Abeta}.


\begin{thebibliography}{10}

\bibitem{Buck1966}
John Buck and Elisabeth Buck.
\newblock Biology of synchronous flashing of fireflies.
\newblock 211:562--564, 1966.

\bibitem{Strogatz2005}
Steven~H Strogatz, Daniel~M Abrams, Allan McRobie, Bruno Eckhardt, and Edward
  Ott.
\newblock {C}rowd synchrony on the millennium bridge.
\newblock {\em Nature}, 438(7064):43--44, 2005.

\bibitem{Wiesenfeld1998}
Kurt Wiesenfeld, Pere Colet, and Steven~H. Strogatz.
\newblock Frequency locking in {J}osephson arrays: {C}onnection with the
  {K}uramoto model.
\newblock {\em Phys. Rev. E}, 57:1563--1569, 1998.

\bibitem{Filatrella2008}
Giovanni Filatrella, Arne~Hejde Nielsen, and Niels~Falsig Pedersen.
\newblock Analysis of a power grid using a kuramoto-like model.
\newblock {\em Eur. Phys. J. B}, 61(4):485--491, 2008.

\bibitem{Kuramoto2003_1}
Yoshiki. Kuramoto.
\newblock {\em Chemical Oscillations, Waves, and Turbulence}.
\newblock Chemistry Series. Dover Publications, 2003.

\bibitem{Peskin1975}
Charles~S Peskin.
\newblock {\em Mathematical aspects of heart physiology}.
\newblock Courant Institute of Mathematical Sciences, New York University New
  York, 1975.

\bibitem{Michaels1987}
Donald~C Michaels, Edward~P Matyas, and Jose Jalife.
\newblock Mechanisms of sinoatrial pacemaker synchronization: a new hypothesis.
\newblock {\em Circ. Res.}, 61:704--714, 1987.

\bibitem{Crook1997}
Sharon~M. Crook, G~Bard Ermentrout, Michael~C. Vanier, and James~M. Bower.
\newblock The role of axonal delay in the synchronization of networks of
  coupled cortical oscillators.
\newblock {\em J. Computat. Neurosci.}, 4(2):161--172, 1997.

\bibitem{Winfree1967}
Arthur~T. Winfree.
\newblock Biological rhythms and the behavior of populations of coupled
  oscillators.
\newblock {\em J. Theor. Biol.}, 16:15--42, 1967.

\bibitem{Kuramoto1975}
Y.~Kuramoto.
\newblock Self-entrainment of a population of coupled non-linear oscillators.
\newblock In H.~Araki, editor, {\em International Symposium on Mathematical
  Problems in Theoretical Physics}, Lecture Notes in Physics, pages 420--422.
  Springer-Verlag Berlin, 1975.

\bibitem{Strogatz2000}
Steven~H Strogatz.
\newblock From {K}uramoto to {C}rawford: {E}xploring the onset of
  synchronization in populations of coupled oscillators.
\newblock {\em Physica D}, 143:1--20, 2000.

\bibitem{Kuramoto2002}
Yoshiki Kuramoto and Dorjsuren Battogtokh.
\newblock Coexistence of coherence and incoherence in nonlocally coupled phase
  oscillators.
\newblock {\em Nonlinear Phenom. Complex Syst.}, 4:380--385, 2002.

\bibitem{Panaggio2014_2}
Mark~J Panaggio and Daniel~M Abrams.
\newblock Chimera states: coexistence of coherence and incoherence in networks
  of coupled oscillators.
\newblock {\em Nonlinearity}, 28(3):R67, 2015.

\bibitem{Abrams2004}
Daniel~M Abrams and Steven~H Strogatz.
\newblock {Chimera states for coupled oscillators}.
\newblock {\em Phys. Rev. Lett.}, 93:174102, 2004.

\bibitem{Shima2004}
Shin-ichiro Shima and Yoshiki Kuramoto.
\newblock Rotating spiral waves with phase-randomized core in nonlocally
  coupled oscillators.
\newblock {\em Phys. Rev. E}, 69:036213, 2004.

\bibitem{Abrams2006}
Daniel~M Abrams and Steven~H Strogatz.
\newblock {Chimera states in a ring of nonlocally coupled oscillators}.
\newblock {\em Int. J. of Bifurc. Chaos}, 16:21--37, 2006.

\bibitem{Abrams2008}
Daniel~M Abrams, Rennie Mirollo, Steven~H Strogatz, and Daniel~A Wiley.
\newblock Solvable model for chimera states of coupled oscillators.
\newblock {\em Phys. Rev. Lett.}, 101:084103, 2008.

\bibitem{Martens2010_1}
Erik~A. Martens, Carlo~R. Laing, and Steven~H. Strogatz.
\newblock Solvable model of spiral wave chimeras.
\newblock {\em Phys. Rev. Lett.}, 104:044101, 2010.

\bibitem{Martens2010_2}
Erik~A. Martens.
\newblock Chimeras in a network of three oscillator populations with varying
  network topology.
\newblock {\em Chaos}, 20(4):043122, 2010.

\bibitem{Omelchenko2012}
Oleh~E Omel'Chenko, Matthias Wolfrum, Serhiy Yanchuk, Yuri~L Maistrenko, and
  Oleksandr Sudakov.
\newblock Stationary patterns of coherence and incoherence in two-dimensional
  arrays of non-locally-coupled phase oscillators.
\newblock {\em Phys. Rev. E}, 85:036210, 2012.

\bibitem{Zhu2012}
Y.~Zhu, Y.~Li, M.~Zhang, and J.~Yang.
\newblock The oscillating two-cluster chimera state in non-locally coupled
  phase oscillators.
\newblock {\em Europhys. Lett.}, 97(1):10009, 2012.

\bibitem{Panaggio2013}
Mark~J. Panaggio and Daniel~M. Abrams.
\newblock Chimera states on a flat torus.
\newblock {\em Phys. Rev. Lett.}, 110:094102, 2013.

\bibitem{Panaggio2014}
Mark~J. Panaggio and Daniel~M. Abrams.
\newblock Chimera states on the surface of a sphere.
\newblock {\em Phys. Rev. E}, 91:022909, 2015.

\bibitem{Laing2009_1}
Carlo~R. Laing.
\newblock Chimera states in heterogeneous networks.
\newblock {\em Chaos}, 19:013113, 2009.

\bibitem{Laing2009_2}
Carlo~R. Laing.
\newblock The dynamics of chimera states in heterogeneous {K}uramoto networks.
\newblock {\em Physica D}, 238(16):1569--1588, 2009.

\bibitem{Laing2012}
Carlo~R. Laing.
\newblock Disorder-induced dynamics in a pair of coupled heterogeneous phase
  oscillator networks.
\newblock {\em Chaos}, 22(4):043104, 2012.

\bibitem{Laing2012_2}
Carlo~R. Laing, Karthikeyan Rajendran, and Ioannis~G. Kevrekidis.
\newblock Chimeras in random non-complete networks of phase oscillators.
\newblock {\em Chaos}, 22(1):013132, 2012.

\bibitem{Yao2013}
Nan Yao, Zi-Gang Huang, Ying-Cheng Lai, and Zhi-Gang Zheng.
\newblock Robustness of chimera states in complex dynamical systems.
\newblock {\em Sci. Rep.}, 3, 2013.

\bibitem{Hagerstrom2012}
Aaron~M Hagerstrom, Thomas~E Murphy, Rajarshi Roy, Philipp H{\"o}vel, Iryna
  Omelchenko, and Eckehard Sch{\"o}ll.
\newblock Experimental observation of chimeras in coupled-map lattices.
\newblock {\em Nature Phys.}, 8:658--661, 2012.

\bibitem{Tinsley2012}
Mark~R. Tinsley, Simbarashe Nkomo, and Kenneth Showalter.
\newblock Chimera and phase-cluster states in populations of coupled chemical
  oscillators.
\newblock {\em Nature Phys.}, 8:662--665, 2012.

\bibitem{Nkomo2013}
Simbarashe Nkomo, Mark Tinsley, and Kenneth Showalter.
\newblock Chimera states in populations of nonlocally coupled chemical
  oscillators.
\newblock {\em Phys. Rev. Lett.}, 110:244102, 2013.

\bibitem{Martens2013}
Erik~Andreas Martens, Shashi Thutupalli, Antoine Fourrière, and Oskar
  Hallatschek.
\newblock Chimera states in mechanical oscillator networks.
\newblock {\em Proc. Nat. Acad. Sci. USA}, 110(4):10563--10567, 2013.

\bibitem{Schmidt2014}
Lennart Schmidt, Konrad Sch\"{o}nleber, Katharina Krischer, and Vladimir
  García-Morales.
\newblock Coexistence of synchrony and incoherence in oscillatory media under
  nonlinear global coupling.
\newblock {\em Chaos}, 24(1):013102, 2014.

\bibitem{Laing2011}
Carlo~R. Laing.
\newblock Fronts and bumps in spatially extended kuramoto networks.
\newblock {\em Physica D}, 240(24):1960--1971, 2011.

\bibitem{Wimmer2014}
Klaus Wimmer, Duane~Q Nykamp, Christos Constantinidis, and Albert Compte.
\newblock Bump attractor dynamics in prefrontal cortex explains behavioral
  precision in spatial working memory.
\newblock {\em Nat. Neurosci.}, 17:431--439, 2014.

\bibitem{Tognoli2014}
Emmanuelle Tognoli and J.~A.~Scott Kelso.
\newblock The metastable brain.
\newblock {\em Neuron}, 81(1):35--48, 2014.

\bibitem{Laing2001}
Carlo~R Laing and Carson~C Chow.
\newblock Stationary bumps in networks of spiking neurons.
\newblock {\em Neural Comput.}, 13(7):1473--1494, 2001.

\bibitem{Ashwin2015}
Peter Ashwin and Oleksandr Burylko.
\newblock Weak chimeras in minimal networks of coupled phase oscillators.
\newblock {\em Chaos}, 25(1):013106, 2015.

\bibitem{wolfrum2015}
Matthias Wolfrum, Oleh Omel'chenko, and Jan Sieber.
\newblock Regular and irregular patterns of self-localized excitation in arrays
  of coupled phase oscillators.
\newblock {\em Chaos}, 25(5):053113, 2015.

\bibitem{Omelchenko2013}
O~E Omel'chenko.
\newblock Coherence--incoherence patterns in a ring of non-locally coupled
  phase oscillators.
\newblock {\em Nonlinearity}, 26(9):2469, 2013.

\bibitem{Wolfrum2011_2}
Matthias Wolfrum and Oleh~E. Omel'chenko.
\newblock Chimera states are chaotic transients.
\newblock {\em Phys. Rev. E}, 84:015201, 2011.

\bibitem{Montbrio2004}
Ernest Montbri\'o, J\"urgen Kurths, and Bernd Blasius.
\newblock Synchronization of two interacting populations of oscillators.
\newblock {\em Phys. Rev. E}, 70:056125, 2004.

\bibitem{Ott2008}
Edward Ott and Thomas~M. Antonsen.
\newblock Low dimensional behavior of large systems of globally coupled
  oscillators.
\newblock {\em Chaos}, 18:037113, 2008.

\bibitem{Ott2009_1}
Edward Ott and Thomas~M. Antonsen.
\newblock Long time evolution of phase oscillator systems.
\newblock {\em Chaos}, 19(2):023117, 2009.

\bibitem{Watanabe1994}
Shinya Watanabe and Steven~H. Strogatz.
\newblock Constants of motion for superconducting {J}osephson arrays.
\newblock {\em Physica D}, 74:197--253, 1994.

\bibitem{Pikovsky2008}
Arkady Pikovsky and Michael Rosenblum.
\newblock Partially integrable dynamics of hierarchical populations of coupled
  oscillators.
\newblock {\em Phys. Rev. Lett.}, 101:264103, 2008.

\bibitem{Laing2010}
Carlo~R. Laing.
\newblock Chimeras in networks of planar oscillators.
\newblock {\em Phys. Rev. E}, 81:066221, 2010.

\bibitem{Kawamura2007}
Yoji Kawamura.
\newblock Chimera ising walls in forced nonlocally coupled oscillators.
\newblock {\em Phys. Rev. E}, 75:056204, 2007.

\bibitem{Sakaguchi2006}
Hidetsugu Sakaguchi.
\newblock Instability of synchronized motion in nonlocally coupled neural
  oscillators.
\newblock {\em Phys. Rev. E}, 73:031907, 2006.

\bibitem{Kuramoto2003_2}
Yoshiki Kuramoto and Shin-ichiro Shima.
\newblock Rotating spirals without phase singularity in reaction-diffusion
  systems.
\newblock {\em Prog. Theor. Phys. Supp.}, 150:115--125, 2003.

\end{thebibliography}
\bibliographystyle{unsrt}

\end{document}